\documentclass[11pt,a4paper]{article}

\usepackage[T1]{fontenc}
\usepackage[utf8]{inputenc}
\usepackage[margin=25mm]{geometry}
\usepackage{amsmath,amssymb,bm}
\usepackage{graphicx}
\usepackage{booktabs}
\usepackage{array}
\usepackage{cite}
\usepackage{microtype}
\usepackage{url}
\usepackage{hyperref}

\hypersetup{
  colorlinks=true,
  linkcolor=blue,
  citecolor=blue,
  urlcolor=blue
}

\title{\textbf{Long-Term Stable Three-Axis Single-Beam Alignment-Based Atomic Magnetometer Using an Uncoated Buffer-Gas-Free Rubidium Ampoule for Unshielded Geomagnetic Monitoring}}

\author{
  Minaru Kawamura\\[6pt]
  \small Department of Electrical and Electronic Engineering, Okayama University of Science,\\
  \small Okayama 700-0005, Japan\\
  \small E-mail: \texttt{minaru@ous.ac.jp}
}

\date{}

\begin{document}

\maketitle

\begin{abstract}
Optically pumped magnetometers achieve exceptional magnetic sensitivity, yet sustaining long-term baseline stability in unshielded Earth-field environments ($\sim 50{,}000~\mathrm{nT}$) remains a major challenge. Here, we present the design, quantum-interference operating principles, calibration, and unshielded validation of a three-axis single-beam alignment-based atomic magnetometer constructed from an unmodified, uncoated, buffer-gas-free commercial rubidium (Rb) glass ampoule and a single 795~nm vertical-cavity surface-emitting laser (VCSEL). Pumped with $45^\circ$-oriented linearly polarized light on the $^{87}\mathrm{Rb}$ $D_1$ ($F=2 \to F'=1$) transition, the sensor exploits ground-state Zeeman quantum interference (rank-2 spin alignment), intrinsically cancelling first-order vector light shifts. To suppress window condensation and thermal magnetic gradients, dual-zone 20~kHz AC heating ($75.0\,^\circ\mathrm{C}$ lower liquid Rb reservoir and $85.0\,^\circ\mathrm{C}$ upper optical zone) is implemented, achieving condensation-free operation over 29~days. Leveraging the absence of buffer-gas pressure broadening, the VCSEL is robustly stabilized via direct absorption spectroscopy at an optimized dither frequency of 3530~Hz, eliminating odd-harmonic intermodulation leakage. A hybrid balanced polarimeter extracts $B_x$, $B_y$, and laser-lock errors from total absorption (sum channel) and $B_z$ from alignment-induced linear dichroism (difference channel) within an active closed-loop zero-field compensation architecture using a 12-bit microcontroller in a dual-stage mid-ranging loop. Spectral analysis demonstrates sub-LSB noise amplitudes of 0.5--5~pT at 1~Hz ($15\text{--}150~\mathrm{pT}/\sqrt{\mathrm{Hz}}$), rolling off to 0.1--1~pT ($3\text{--}30~\mathrm{pT}/\sqrt{\mathrm{Hz}}$) near 10~Hz. Continuous unshielded geomagnetic monitoring in Okayama benchmarked against synthesized Kakioka and Kanoya observatory records demonstrates high vector tracking fidelity and establishes an upper bound on instrumental bias instability of $\le 0.1~\mathrm{nT}$ at integration times of 100--200~s.
\end{abstract}


\section{Introduction}
\label{sec:intro}

Optically pumped magnetometers (OPMs) utilizing alkali-metal atomic vapor have emerged as leading quantum magnetic sensors, offering sub-picotesla to sub-femtotesla sensitivities without the cryogenic infrastructure required by superconducting quantum interference devices (SQUIDs)~\cite{budker2007,kominis2003}. While spin-exchange relaxation-free (SERF) magnetometers achieve record sensitivities inside multi-layer magnetic shields, operating compact three-axis vector atomic magnetometers continuously in unshielded Earth-field environments ($\sim 50{,}000~\mathrm{nT}$) remains a critical requirement for practical applications in geomagnetic observatory networks, geophysical surveys, and mobile biomagnetic sensing~\cite{boto2018,holmes2019,seltzer2004}.

Most conventional vector OPMs rely on circularly polarized light to generate rank-1 atomic spin orientation (macroscopic dipole magnetization)~\cite{budker2007,seltzer2004}. However, circularly polarized pumping inherently induces a first-order vector AC Stark shift (light shift), which acts as a fictitious magnetic field along the optical axis proportional to laser detuning and intensity fluctuations, severely limiting long-term baseline stability~\cite{mathur1968,weis2006}. In contrast, atomic magnetometers pumped with linearly polarized light operate via quantum mechanical interference among ground-state Zeeman sublevels---specifically, preparing rank-2 quadrupole spin alignment ($\Delta m = \pm 2$ Zeeman coherences and coherent dark states) probed through nonlinear magneto-optical rotation (NMOR) and resonant absorption~\cite{mathur1968,weis2006,budker2002,meraki2023}. Because linearly polarized light consists of an equal coherent superposition of $\sigma^+$ and $\sigma^-$ components, the first-order vector light shifts cancel intrinsically, leaving only higher-order tensor shifts that do not mimic a Zeeman magnetic field~\cite{weis2006}. Furthermore, operating in an alignment-based regime with high-frequency field modulation or harmonic (such as second-harmonic, $2f$) phase-sensitive detection allows the magnetic signal to be shifted well above the low-frequency $1/f$ technical flicker noise of semiconductor diode lasers~\cite{weis2006,dupontroc1969,legal2022}.

Despite these fundamental advantages, deploying a single-beam three-axis alignment-based magnetometer in an unshielded environment poses three interrelated experimental challenges: thermal laser wavelength drift, atomic density stability without complex cell fabrication, and inter-axis coupling~\cite{budker2007,huang2016}. Vertical-cavity surface-emitting lasers (VCSELs), favored for low power consumption and compactness, exhibit steep thermal wavelength tuning coefficients ($\sim 0.06~\mathrm{nm}/^\circ\mathrm{C}$)~\cite{affolderbach2005}, while sub-Doppler locking methods such as saturated absorption spectroscopy (SAS) add counter-propagating optical paths that can introduce mechanical instabilities in field-deployable instruments~\cite{gruet2013}. Moreover, in a single-beam linearly polarized configuration, the geometric orientation of the optical polarization axis introduces inherent cross-axis interactions between transverse magnetic field components (particularly between the $y$- and $z$-axes when the incident polarization is rotated by $45^\circ$ for balanced polarimetry), which must be carefully isolated under dynamic ambient disturbances.

In this paper, we present the design, quantum-interference operating principles, calibration, and unshielded geomagnetic validation of a long-term stable, three-axis single-beam alignment-based atomic magnetometer constructed from an unmodified, uncoated, buffer-gas-free commercial rubidium (Rb) glass ampoule and a single 795~nm VCSEL (Fig.~\ref{fig:setup}). Taking advantage of the absence of buffer-gas pressure broadening, the VCSEL frequency is robustly stabilized via direct absorption spectroscopy (DAS) across a wide capture range, while the relatively broad transit-time and wall-collision relaxation bandwidth ($\sim\mathrm{kHz}$) enables high-frequency orthogonal field modulation (1.29--5.46~kHz) well above low-frequency laser flicker noise without severe response roll-off. To extract all three magnetic vector components simultaneously while suppressing inter-channel crowding and inter-axis coupling, we implement a hybrid balanced-polarimeter readout combining total absorption (sum channel for $B_x$, $B_y$, and laser locking) with alignment-induced magneto-optical birefringence/dichroism (difference channel for $B_z$), embedded inside an active closed-loop zero-field compensation architecture with dual-zone 20~kHz AC heating. Using low-cost commercial electronics---analog phase-sensitive detectors (AD630) and a 12-bit microcontroller (NI myRIO) in a dual-stage mid-ranging loop---we demonstrate sub-LSB spectral noise resolution, characterize both static and dynamic ($Z \to Y$) inter-axis interactions, and validate multi-day unshielded geomagnetic tracking in Okayama against a synthesized reference from the Kakioka and Kanoya Magnetic Observatories.

\section{Theoretical Framework: Quantum Interference and Three-Axis Vector Extraction}
\label{sec:theory}

\subsection{Density Matrix Formalism and Ground-State Zeeman Coherence}
\label{sec:formalism}

Unlike circularly polarized magnetometers that generate rank-1 dipole spin orientation ($\langle \hat{\bm{F}} \rangle \neq 0$), our single-beam instrument operates with linearly polarized light tuned to the $^{87}\mathrm{Rb}$ $D_1$ ($F=2 \to F'=1$) transition, probing quantum mechanical interference among ground-state Zeeman sublevels~\cite{weis2006,budker2002,meraki2023,happer1972,auzinsh2010}. In an $F \to F-1$ hyperfine transition, coherent two-photon ($\Lambda$-type) excitation by linearly polarized light---an equal-amplitude coherent superposition of $\sigma^+$ and $\sigma^-$ components---prepares the atomic ensemble in coherent population trapping (CPT) dark states that are decoupled from the optical field~\cite{budker2002,auzinsh2010,fleischhauer2005}. Because the $\sigma^+$ and $\sigma^-$ components carry equal and opposite helicities, first-order vector AC Stark shifts cancel intrinsically, eliminating fictitious magnetic field drift along the optical axis~\cite{mathur1968,weis2006}.

The evolution of the ground-state atomic density operator $\hat{\rho}$ in an external magnetic field $\bm{B}$ is governed by the Liouville--von Neumann master equation~\cite{happer1972,auzinsh2010}:
\begin{equation}
  \frac{d\hat{\rho}}{dt} = -\frac{i}{\hbar}\left[\hat{H}_B, \hat{\rho}\right] + \hat{\mathcal{R}}_{\mathrm{op}}(\hat{\rho}) - \hat{\Gamma}(\hat{\rho}),
  \label{eq:liouville}
\end{equation}
where $\hat{H}_B = g_F \mu_B \bm{B} \cdot \hat{\bm{F}} = \hbar \bm{\Omega}_L \cdot \hat{\bm{F}}$ is the Zeeman Hamiltonian with Larmor frequency vector $\bm{\Omega}_L = \gamma \bm{B}$ ($\gamma/2\pi \approx 7~\mathrm{Hz/nT}$ for $^{87}\mathrm{Rb}$), $\hat{\mathcal{R}}_{\mathrm{op}}$ is the optical pumping superoperator, and $\hat{\Gamma}$ represents spin relaxation. Expanding $\hat{\rho}$ in irreducible spherical tensor operators $\hat{T}^{(k)}_q$ ($\hat{\rho} = \sum_{k,q} \rho^{(k)}_q \hat{T}^{(k)\dagger}_q$), linearly polarized pumping selectively prepares the rank-2 quadrupole alignment multipole components $\rho^{(2)}_q$ ($q = 0, \pm 1, \pm 2$, corresponding to $\Delta m = 0, \pm 1, \pm 2$ Zeeman coherences) while maintaining zero rank-1 orientation ($k=1$)~\cite{weis2006,budker2002,auzinsh2010}.

In contrast to alignment-based magnetometers employing high-pressure buffer-gas cells to suppress wall diffusion~\cite{rushton2023}, the unmodified commercial ampoule used in this study operates in a buffer-gas-free, high-Knudsen-number ballistic regime. Consequently, the effective relaxation rate $\Gamma$ is governed by wall collisions and atomic transit time across the 3~mm laser beam ($\Gamma / 2\pi \sim \mathrm{kHz}$). Although this yields a broader magnetic resonance than in buffer-gas or paraffin-coated cells, the absence of collisional pressure broadening resolves the 814.5~MHz excited-state hyperfine splitting within the Doppler profile for simple direct absorption locking, while the kHz-scale relaxation bandwidth allows high-frequency orthogonal field modulation (1.29--5.46~kHz) well above the $1/f$ laser flicker noise regime without severe response roll-off. Furthermore, maintaining $|\bm{B}| \approx 0$ ($\bm{\Omega}_L \approx 0$) via closed-loop compensation ensures that the 20~kHz AC cell heater acts as a completely adiabatic perturbation whose residual spatial gradients time-average to zero.

\subsection{Orthogonal Field Modulation, Hybrid Polarimeter Readout, and Inter-Axis Coupling}
\label{sec:hybrid}

In our optical arrangement (Fig.~\ref{fig:setup}), the 795~nm beam propagates along the $x$-axis with its linear polarization rotated by $45^\circ$ in the $y$--$z$ plane ($\hat{\bm{\epsilon}} = (\hat{\bm{y}} + \hat{\bm{z}})/\sqrt{2}$) via a half-wave plate upstream of the vapor cell. Downstream of the cell, a polarizing beam splitter (PBS) projects the transmitted light onto orthogonal horizontal ($\hat{\bm{y}}$, $\mathrm{PD}_1$) and vertical ($\hat{\bm{z}}$, $\mathrm{PD}_2$) polarizations. To simultaneously extract all three magnetic field components from a single beam~\cite{huang2016,xiao2021}, orthogonal sinusoidal modulation fields are applied via three Helmholtz coils:
\begin{equation}
  B_x(t) = B_{x,\mathrm{DC}} + A_x \sin(\omega_x t), \quad
  B_y(t) = B_{y,\mathrm{DC}} + A_y \cos(\omega_y t), \quad
  B_z(t) = B_{z,\mathrm{DC}} + A_z \sin(\omega_z t).
  \label{eq:mod}
\end{equation}

The symmetric traceless Cartesian alignment tensor $A_{ij}$ representing $\rho^{(2)}_q$ obeys the steady-state precession balance $\Gamma (A_{ij} - A^{(0)}_{ij}) = [\bm{\Omega}_L \times, A]_{ij}$, where the pumped zero-field alignment is $A^{(0)}_{ij} \propto \epsilon_i \epsilon_j - \frac{1}{3}\delta_{ij}$~\cite{weis2006,auzinsh2010}. Solving this equation perturbatively in the near-zero-field regime ($|\bm{\Omega}_L| \ll \Gamma$) yields the two balanced-polarimeter readout observables (Fig.~\ref{fig:setup}):

\begin{enumerate}
  \item \textbf{Sum Channel ($I_{\mathrm{Sum}} = I_{\mathrm{PD1}} + I_{\mathrm{PD2}}$, Total Absorption via Instrumentation Amplifier):}
  Total optical transmission probes $\hat{\bm{\epsilon}}^T A \hat{\bm{\epsilon}}$, reflecting the disruption of dark-state CPT coherence by magnetic field components orthogonal to $\hat{\bm{\epsilon}}$ ($|\bm{\Omega}_L \times \hat{\bm{\epsilon}}|^2 = \Omega_x^2 + \frac{1}{2}(\Omega_y - \Omega_z)^2$):
  \begin{equation}
    \Delta I_{\mathrm{Sum}}(\bm{\Omega}_L) \propto -\frac{2}{\Gamma^2}\left[\Omega_x^2 + \frac{1}{2}(\Omega_y - \Omega_z)^2\right] + O\!\left(\frac{\Omega_L^4}{\Gamma^4}\right).
    \label{eq:sum}
  \end{equation}
  First-harmonic ($1\omega_i$) phase-sensitive demodulation yields the partial derivatives $\partial \Delta I_{\mathrm{Sum}} / \partial B_i$ around null field:
  \begin{equation}
    V_x \propto -\frac{4\gamma}{\Gamma^2}\,\Omega_{x,\mathrm{DC}}, \qquad
    V_y \propto -\frac{2\gamma}{\Gamma^2}\,(\Omega_{y,\mathrm{DC}} - \Omega_{z,\mathrm{DC}}).
    \label{eq:vxvy}
  \end{equation}
  Two crucial physical consequences follow directly from these expressions. First, because a magnetic field strictly parallel to $\hat{\bm{\epsilon}}$ ($\Omega_y = \Omega_z, \Omega_x = 0$) commutes with the pumped alignment state and causes zero precession, $\Delta I_{\mathrm{Sum}}$ is completely invariant to $(\Omega_y + \Omega_z)$. Consequently, attempting to extract both $B_y$ and $B_z$ from total absorption alone yields collinear, linearly dependent error signals ($\partial \Delta I_{\mathrm{Sum}}/\partial B_z = -\partial \Delta I_{\mathrm{Sum}}/\partial B_y$), making full three-axis vector reconstruction mathematically impossible from the sum channel alone. Second, along the propagation axis ($x$-axis), velocity-selective CPT saturation of slow atoms in the buffer-gas-free vapor produces a sharp sub-linewidth transparency dip governed by $\Delta m_x = 2$ precession ($2\Omega_x$, half-width $\Gamma/2$), yielding the steep difference-of-two-Lorentzians curve for $B_x$ in Fig.~\ref{fig:sweep}. In contrast, the transverse $B_y$ absorption response mixes $\Delta m = 2$ and broader $\Delta m = 1$ ($\Omega_y$, half-width $\Gamma$) coherences with an additional geometric factor of $\sin^2 45^\circ = 1/2$, explaining why the $B_y$ dispersion curve in Fig.~\ref{fig:sweep} is roughly twice as wide with approximately half the central slope of $B_x$.

  \item \textbf{Difference Channel ($I_{\mathrm{Diff}} = I_{\mathrm{PD1}} - I_{\mathrm{PD2}}$, Alignment-Induced Linear Dichroism via Differential Transimpedance Amplifier):}
  Subtracting the orthogonal PBS photocurrents directly at the transimpedance input probes $A_{yy} - A_{zz}$---the alignment-induced linear dichroism (and nonlinear magneto-optical rotation, NMOR) between the $\hat{\bm{y}}$ and $\hat{\bm{z}}$ axes~\cite{budker2002,auzinsh2010}:
  \begin{equation}
    \Delta I_{\mathrm{Diff}}(\bm{\Omega}_L) \propto -\frac{2\Omega_x}{\Gamma} + \frac{\Omega_y^2 - \Omega_z^2}{\Gamma^2} + O\!\left(\frac{\Omega_L^3}{\Gamma^3}\right).
    \label{eq:diff}
  \end{equation}
  Physically, taking $\hat{\bm{z}}$ as the quantization axis, the $\hat{\bm{z}}$-polarized PBS output ($I_{\mathrm{PD2}}$) drives only $\Delta m_z = 0$ ($\pi$) transitions that are insensitive to $B_z$, whereas the $\hat{\bm{y}}$-polarized output ($I_{\mathrm{PD1}}$) drives coherent $\Delta m_z = \pm 1$ ($\sigma^\pm$) transitions whose $\Delta m_z = 2$ dark-state coherence dephases at $2\Omega_z$, selectively increasing absorption on $\mathrm{PD}_1$. Crucially, the cross term $\Omega_y \Omega_z$ cancels identically between $A_{yy}$ and $A_{zz}$. Demodulating $I_{\mathrm{Diff}}$ at $1\omega_z$ therefore isolates the vertical field error:
  \begin{equation}
    V_z \propto \frac{\partial \Delta I_{\mathrm{Diff}}}{\partial B_z} \propto -\frac{2\gamma}{\Gamma^2}\,\Omega_{z,\mathrm{DC}},
    \label{eq:vz}
  \end{equation}
  with the narrow $\Delta m_z = 2$ linewidth ($\Gamma/2$, matching $B_x$ in Fig.~\ref{fig:sweep}) while simultaneously rejecting quiescent DC photocurrents and common-mode laser intensity noise (RIN) prior to amplification.
\end{enumerate}

Combining the three demodulated error signals yields a full-rank, upper-triangular linear sensitivity Jacobian around zero field:
\begin{equation}
  \begin{pmatrix} V_x \\ V_y \\ V_z \end{pmatrix}
  \propto
  \begin{pmatrix}
    -4 & 0 & 0 \\
    0 & -2 & +2 \\
    0 & 0 & -2
  \end{pmatrix}
  \begin{pmatrix} B_{x,\mathrm{DC}} \\ B_{y,\mathrm{DC}} \\ B_{z,\mathrm{DC}} \end{pmatrix}.
  \label{eq:jacobian}
\end{equation}
This structure rigorously explains the observed inter-axis interactions: $B_x$ is completely decoupled to first order, and $B_z$ (from the balanced difference channel) has zero first-order additive sensitivity to $B_x$ or $B_y$ perturbations. Conversely, the $B_y$ absorption channel possesses an inherent 1:1 (100\%) additive coupling to uncompensated $B_z$ fields. Under quasi-DC closed-loop operation, the $B_z$ feedback loop continuously nulls $B_{z,\mathrm{DC}} \to 0$, causing this $Z \to Y$ coupling to vanish statically. However, during rapid dynamic disturbances (such as 0.1~Hz calibration fields or steep railway spikes), any transient tracking lag in the $B_z$ integral loop leaves a $90^\circ$ out-of-phase residual field $\delta B_z \propto j\omega B_{z,\mathrm{ext}}$ inside the cell, which directly injects an imaginary (derivative) cross-talk component into $B_y$, as confirmed experimentally in Secs.~\ref{sec:calib} and~\ref{sec:step}.

\subsection{Selection of Non-Harmonic Modulation Frequencies}
\label{sec:freq}

Because the analog balanced demodulators (AD630) employ square-wave switching multiplication, each channel responds not only to its fundamental reference frequency $f_i = \omega_i / 2\pi$ but also to its odd harmonics $(2k+1)f_i$~\cite{ad630}. To eliminate both odd-harmonic cross-coupling and low-order intermodulation distortion within the 10~Hz demodulation low-pass filter (LPF), the modulation frequencies were optimized to mutually non-harmonic values: $f_x = 5460~\mathrm{Hz}$, $f_y = 1290~\mathrm{Hz}$, and $f_z = 2130~\mathrm{Hz}$ for the three magnetic axes, with the VCSEL direct-absorption dither set to $f_{\mathrm{lock}} = 3530~\mathrm{Hz}$ (at a reduced modulation amplitude of one-tenth that of the magnetic channels).

In earlier preliminary configurations using $f_{\mathrm{lock}} = 3220~\mathrm{Hz}$ and $f_x = 5770~\mathrm{Hz}$, the second harmonic of the laser dither ($2f_{\mathrm{lock}} = 6440~\mathrm{Hz}$) fell within 10~Hz of $5f_y = 6450~\mathrm{Hz}$, and $f_{\mathrm{lock}} + 2f_y = 5800~\mathrm{Hz}$ lay within 30~Hz of $f_x = 5770~\mathrm{Hz}$. Furthermore, shifting only $f_{\mathrm{lock}}$ to 3530~Hz placed it at the exact midpoint between 1290~Hz and 5770~Hz, creating a zero-frequency third-order intermodulation overlap ($2f_{\mathrm{lock}} - f_x = f_y$). By additionally shifting $f_x$ from 5770~Hz to 5460~Hz, every second-, third-, and fourth-order intermodulation product among the three sum-channel frequencies ($f_y$, $f_{\mathrm{lock}}$, and $f_x$) is separated by at least 300~Hz from all demodulation carriers (closest product: $4f_y = 5160~\mathrm{Hz}$ vs.\ $f_x = 5460~\mathrm{Hz}$, followed by third-order terms spaced by 310~Hz and $f_{\mathrm{lock}} + 2f_y = 6110~\mathrm{Hz}$ spaced by 650~Hz), while the nearest AD630 odd-harmonic overlap with the 20~kHz AC cell heater ($15f_y = 19{,}350~\mathrm{Hz}$) lies 650~Hz away. Consequently, all intermodulation and switching-harmonic products are attenuated by more than 59~dB by the second-order 10~Hz LPFs, cleanly revealing the intrinsic sub-picotesla noise floor of the magnetometer.

\section{Active Magnetic Field Compensation and Dual-Stage Mid-Ranging Control}
\label{sec:comp}

Because the high-sensitivity zero-field dispersive resonances have a narrow linear dynamic range of approximately $\pm 200~\mathrm{nT}$ ($\pm 2~\mathrm{mGauss}$, Fig.~\ref{fig:sweep}), continuous unshielded operation under the ambient geomagnetic field ($\sim 50{,}000~\mathrm{nT}$) requires active magnetic field compensation~\cite{holmes2019,seltzer2004,yan2022,zhang2022,zhang2020}. In our three-axis closed-loop architecture (Fig.~\ref{fig:setup}), demodulated error signals from the phase-sensitive detectors are fed to independent proportional--integral--derivative (PID) controllers that drive constant current sources (CCS) connected to orthogonal feedback coils, continuously nulling the total field inside the vapor cell to $|\bm{B}| \approx 0$. Beyond expanding the dynamic range to the compliance limits of the current drivers, active zero-field compensation serves three essential physical functions in this alignment-based magnetometer: (i) it continuously enforces $B_{z,\mathrm{DC}} \to 0$, statically eliminating the intrinsic 1:1 $Z \to Y$ additive cross-coupling derived in Sec.~\ref{sec:hybrid}; (ii) it maintains $\bm{\Omega}_L \approx 0$ so that the 20~kHz AC cell heating remains strictly adiabatic; and (iii) it suppresses scale-factor drift caused by slow variations in atomic vapor density or optical power, since the external field is read directly from the coil compensation currents.

A central instrumentation challenge in unshielded closed-loop magnetometry is reconciling a wide static-plus-diurnal compensation range ($\sim 50{,}000~\mathrm{nT}$) with picotesla-level digital resolution without resorting to expensive 24-bit converters or custom FPGA hardware. Directly spanning $\pm 50{,}000~\mathrm{nT}$ with a 12-bit digital-to-analog converter (DAC) would yield an unacceptably coarse quantization step of 24.4~nT/LSB. To overcome this trade-off, we implement a dual-stage mid-ranging control strategy~\cite{allison1998} (Fig.~\ref{fig:setup}), partitioning the total compensated magnetic field on each axis into a slow main-bias stage and a fast high-resolution stage:
\begin{equation}
  B_{i,\mathrm{total}}(t) = B_{i,\mathrm{CCS}}(t) + B_{i,\mathrm{DAC}}(t) \qquad (i = x, y, z).
  \label{eq:dual}
\end{equation}
In this architecture, the digital PID controller outputs a fast feedback signal through a 12-bit DAC spanning a narrow $\pm 100~\mathrm{nT}$ compensation window, yielding a discrete quantization step of $\Delta = 200~\mathrm{nT}/4096 \approx 48.8~\mathrm{pT/LSB}$. Concurrently, an analog proportional--integral (PI) controller monitors the 12-bit DAC output and slowly adjusts the programmable main CCS to offload static background fields and slow diurnal drifts, keeping the DAC operating point continuously centered near zero. Consequently, the DAC never saturates or suffers integrator windup during large diurnal excursions or abrupt railway disturbances. For full-bandwidth geomagnetic recording (Secs.~\ref{sec:obs} and~\ref{sec:allan}), both stages are summed to reconstruct $B_{i,\mathrm{total}}(t)$; meanwhile, above the low-frequency mid-ranging crossover ($f_{\mathrm{PI}} \sim 10\text{--}30~\mathrm{mHz}$), rapid fluctuations are tracked directly by the fast DAC stage $B_{i,\mathrm{DAC}}(t)$.

For a 12-bit quantizer with step size $\Delta = 48.8~\mathrm{pT}$, the theoretical full-band root-mean-square (RMS) quantization noise is $\sigma_q = \Delta / \sqrt{12} \approx 14.1~\mathrm{pT}$. When updated deterministically at $f_s = 250~\mathrm{Hz}$ (Nyquist frequency $f_N = 125~\mathrm{Hz}$), this corresponds to a uniform quantization noise spectral density of $\sigma_q / \sqrt{f_N} \approx 1.26~\mathrm{pT}/\sqrt{\mathrm{Hz}}$. Crucially, the inherent broadband analog noise of the sensor---comprising optical shot noise, transimpedance Johnson noise, and residual laser intensity fluctuations---acts as a natural, zero-mean Gaussian additive stochastic dither signal prior to digitization~\cite{carbone1994}. By continuously oversampling this dithered error signal at 250~Hz and filtering within the closed-loop detection bandwidth, the effective number of bits (ENOB) is enhanced, seamlessly recovering sub-LSB magnetic resolution (down to 0.3--2~pT FFT amplitude at 1~Hz, as verified in Sec.~\ref{sec:noise}).

Finally, closed-loop stability requires careful matching of the digital PID gain to the phase response of the analog second-order 10~Hz demodulation LPF. Excessively high PID gain pushes the unity-gain crossover frequency into the phase-lag regime of the 10~Hz LPF, reducing phase margin and inducing closed-loop sensitivity peaking, DAC chattering, and amplified transient $Z \to Y$ derivative overshoots. Properly damping the loop gain preserves unity-gain tracking ($|T(f)| \approx 1$) across the geomagnetic signal band while eliminating self-excited high-frequency noise.

\section{Direct Absorption Frequency Locking of the VCSEL}
\label{sec:das}

Laser wavelength stability is vital for sustaining sub-nanotesla baseline accuracy over multi-day unshielded operation. Standard 795~nm VCSELs exhibit steep thermal tuning coefficients of approximately $0.06~\mathrm{nm}/^\circ\mathrm{C}$ ($\sim 28.5~\mathrm{GHz}/^\circ\mathrm{C}$ at 795~nm)~\cite{affolderbach2005}, meaning that an ambient temperature drift of just $0.02\,^\circ\mathrm{C}$ shifts the emission frequency across the entire Doppler-broadened atomic absorption line. Although linearly polarized pumping intrinsically cancels first-order vector light shifts (Sec.~\ref{sec:formalism})~\cite{weis2006}, unstabilized VCSEL detuning still degrades magnetometer stability through three mechanisms: (i) severe FM-to-AM noise conversion on the steep slopes of the absorption line, injecting laser frequency noise directly into the sum-channel ($B_x, B_y$) readout; (ii) modulation of the optical pumping rate $R_{\mathrm{op}}$ and resonance scale factor; and (iii) higher-order tensor AC Stark shifts~\cite{mathur1968,weis2006} as well as minor residual vector light shifts arising from slight stress-induced birefringence in the unmodified glass ampoule wall.

While sub-Doppler techniques such as saturated absorption spectroscopy (SAS) provide megahertz-wide frequency references~\cite{macadam1992}, their counter-propagating beam paths and additional beam-splitting optics increase mechanical complexity and narrow the lock capture range, rendering them susceptible to lock dropout under thermal or acoustic shocks in field deployments~\cite{gruet2013}. In contrast, our buffer-gas-free rubidium ampoule exhibits negligible collisional pressure broadening. For the $^{87}\mathrm{Rb}$ $D_1$ line ($5^2S_{1/2} \to 5^2P_{1/2}$, 794.979~nm), the excited-state hyperfine splitting between $F'=1$ and $F'=2$ is 814.5~MHz~\cite{steck2024}. At the upper optical-zone temperature of $T_2 = 85.0\,^\circ\mathrm{C}$ (358.15~K), the Doppler full width at half maximum (FWHM) is
\begin{equation}
  \Delta \nu_D = \nu_0 \sqrt{\frac{8 k_B T_2 \ln 2}{m c^2}} \approx 548~\mathrm{MHz},
  \label{eq:doppler}
\end{equation}
where $\nu_0 = 377.107~\mathrm{THz}$ and $m$ is the $^{87}\mathrm{Rb}$ atomic mass~\cite{steck2024}. Because $\Delta \nu_D$ is well below the 814.5~MHz hyperfine splitting, the $^{87}\mathrm{Rb}$ $F=2 \to F'=1$ absorption peak is clearly resolved in simple single-pass transmission without sub-Doppler optics.

We therefore implement active frequency locking via direct absorption spectroscopy (DAS) using the primary sensing beam~\cite{affolderbach2005,affolderbach2000,kitching2018} (Fig.~\ref{fig:setup}). A small sinusoidal current dither at $f_{\mathrm{lock}} = 3530~\mathrm{Hz}$ (with a modulation amplitude set to one-tenth of the magnetic modulation drives to avoid parasitic broadening and intensity modulation) is superimposed onto the VCSEL DC injection current. Demodulating the sum-channel output ($I_{\mathrm{PD1}} + I_{\mathrm{PD2}}$) at 3530~Hz generates the first derivative of the Doppler-broadened $^{87}\mathrm{Rb}$ $F=2 \to F'=1$ absorption profile, producing a dispersion error signal with a linear zero-crossing precisely at the absorption maximum---where FM-to-AM conversion also vanishes to first order. Feeding this error signal back to the VCSEL injection current source (combined with precision laser thermoelectric temperature control) locks the emission wavelength continuously. The $\sim 550~\mathrm{MHz}$ Doppler linewidth provides a wide ($\sim\mathrm{GHz}$) capture range that prevents lock dropout during abrupt environmental transients, while suppressing $1/f$ wavelength drift over multi-week operation~\cite{affolderbach2005,kitching2018}. Furthermore, because the blue wing of the $^{87}\mathrm{Rb}$ $F=2 \to F'=1$ line exhibits slight asymmetric overlap with neighboring $F=2 \to F'=2$ and $^{85}\mathrm{Rb}$ Doppler wings, maintaining millikelvin-level cell temperature regulation (Sec.~\ref{sec:cell}) stabilizes the Doppler width $\Delta \nu_D$ and prevents thermal asymmetry from shifting the DAS lock zero-crossing point.

\section{Experimental Setup}
\label{sec:setup}

\begin{figure}[htbp]
  \centering
  \includegraphics[width=0.95\linewidth]{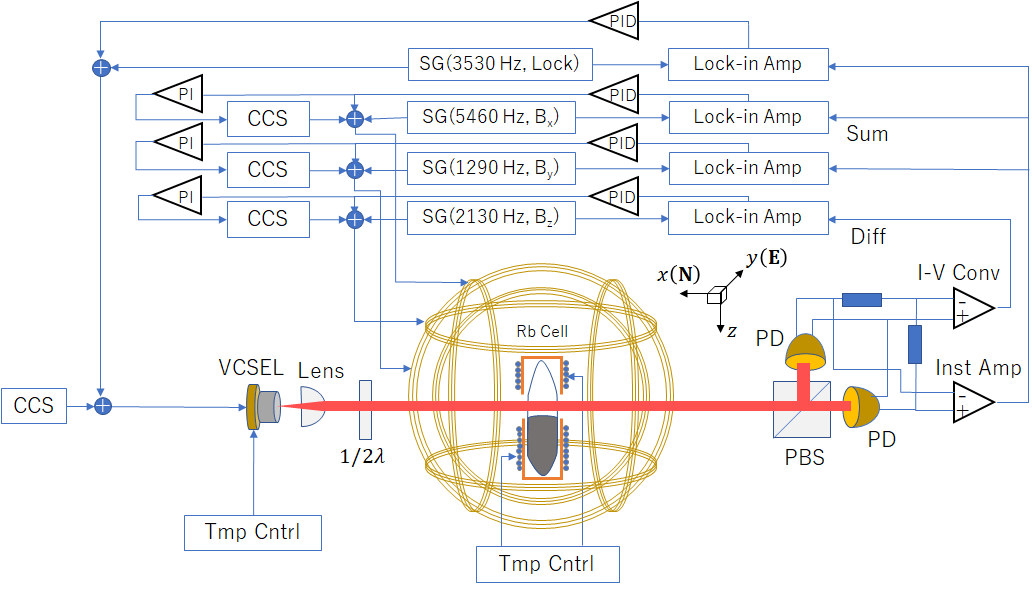}
  \caption{Schematic diagram of the three-axis closed-loop single-beam alignment-based atomic magnetometer. A collimated 795~nm beam from a VCSEL passes through a half-wave plate ($\lambda/2$ at $45^\circ$ positioned immediately upstream of the cell) and enters an unmodified commercial Rb glass ampoule containing 1~g of liquid rubidium metal, housed in a four-slit copper sleeve wrapped in a 5-mm fluororubber jacket and heated by a dual-zone ($75.0\,^\circ\mathrm{C}$ lower / $85.0\,^\circ\mathrm{C}$ upper) 20~kHz AC non-inductive heater. Downstream of the cell, a polarizing beam splitter (PBS) and balanced photodiodes (PDs) split the transmitted light. The sum channel, amplified via an instrumentation amplifier (\texttt{Inst Amp}), provides total optical absorption signals for direct-absorption laser frequency locking (3530~Hz) and the $B_x$ (5460~Hz) and $B_y$ (1290~Hz) field error signals. The difference channel, detected directly via a differential transimpedance amplifier (\texttt{I-V Conv}), measures alignment-induced linear dichroism for the $B_z$ field error signal (2130~Hz). Four AD630 phase-sensitive detectors (10~Hz second-order LPF) and an embedded NI myRIO controller (4~ms loop) drive three-axis 5~cm feedback coils via dual-stage analog PI mid-ranging controllers and constant current sources (CCS) to maintain active zero-field compensation.}
  \label{fig:setup}
\end{figure}

The sensor head and control instrumentation are engineered to minimize size, weight, power, and cost (SWaP-C) while maximizing long-term unshielded stability~\cite{kitching2018}. The overall schematic of the three-axis closed-loop single-beam alignment-based magnetometer is illustrated in Fig.~\ref{fig:setup}.

\subsection{Sensor Cell and Dual-Zone 20 kHz Non-Magnetic Heating Mechanism}
\label{sec:cell}

The atomic medium is an unmodified, vacuum-sealed commercial natural rubidium glass ampoule (Sigma-Aldrich, outer diameter 7~mm, total length 50~mm). The lower 30~mm contains approximately 1~g of rubidium metal, leaving the upper 20~mm as the active optical vapor space (Fig.~\ref{fig:setup}). Because the operating temperature exceeds the melting point of rubidium ($39.3\,^\circ\mathrm{C}$)~\cite{steck2024}, the rubidium resides as a 1~g liquid metallic pool at the bottom of the ampoule. This large reactive metal reservoir serves as a continuous chemical getter, irreversibly binding reactive trace gases (such as oxygen and water vapor) that outgas from or permeate the glass envelope over extended deployment periods, thereby preserving internal vacuum quality and stable relaxation rates.

To prevent rubidium condensation on the upper optical path while maintaining a precisely regulated atomic vapor density, we implemented a dual-zone thermal architecture inside a cylindrical copper sleeve wrapped externally with a 5-mm-thick fluororubber thermal insulation jacket (Fig.~\ref{fig:setup}). Four longitudinal slits are machined along the copper cylinder to interrupt circumferential eddy-current paths and prevent attenuation or phase lag of the kHz AC magnetic modulation fields. Two independent non-inductive heating elements---fabricated from twisted 0.05~mm enameled copper wire wound in a four-fold folded geometry---surround the lower reservoir zone and the upper optical zone, respectively. Under closed-loop thermal PID control, the lower liquid reservoir is regulated at $T_1 = 75.000 \pm 0.007\,^\circ\mathrm{C}$ to set the saturated rubidium vapor pressure, while the upper optical zone is held at $T_2 = 85.000 \pm 0.017\,^\circ\mathrm{C}$. This $+10\,^\circ\mathrm{C}$ vertical temperature gradient thermodynamically suppresses alkali condensation on the optical path: continuous operation exceeding 29~days ($> 2.5 \times 10^6~\mathrm{s}$) exhibited no measurable degradation in transmitted optical power.

Although non-inductive winding cancels most of the heater magnetic field, DC electrical heating still produces residual spatial field gradients across the nearby vapor space that broaden the zero-field resonance and introduce low-frequency magnetic drift~\cite{lu2020}. Both heater zones are therefore driven by a 20~kHz AC current coupled through series DC-blocking capacitors to eliminate amplifier DC offset leakage and low-frequency thermal-controller switching transients~\cite{lu2020}. Selecting 20~kHz avoids the excessive inductive reactance voltage drop ($\omega L$) encountered in the long twisted windings at higher frequencies (e.g., 70~kHz)---providing sufficient thermal power headroom to sustain the $85.0\,^\circ\mathrm{C}$ upper zone across the $10\,^\circ\mathrm{C}$ inter-zone gradient---while remaining well above the atomic relaxation rate, all modulation carriers (1.29--5.46~kHz), and the 10~Hz demodulation filter cutoff (attenuated by $-132~\mathrm{dB}$). Comparative spectral tests between 20~kHz and 40~kHz heater excitation confirmed identical magnetic noise performance.

\subsection{Optical Train and Hybrid Polarimeter Readout Architecture}
\label{sec:optics}

As shown in Fig.~\ref{fig:setup}, a single beam emitted from a 795~nm VCSEL tuned to the $^{87}\mathrm{Rb}$ $D_1$ line is collimated by an aspheric lens to a beam diameter of 3~mm. Immediately after the collimator---upstream of the vapor cell---a half-wave plate ($\lambda/2$) rotates the incident linear polarization by $45^\circ$ relative to the optical bench ($y$--$z$ plane) before the beam traverses the upper optical space of the Rb ampoule along the $x$-axis. Downstream of the cell, a polarizing beam splitter (PBS) resolves the transmitted beam into orthogonal horizontal ($\hat{\bm{y}}$) and vertical ($\hat{\bm{z}}$) linear polarizations detected by a balanced pair of photodiodes (PDs).

Because the buffer-gas-free cell exhibits no collisional pressure broadening, the 814.5~MHz excited-state hyperfine splitting of $^{87}\mathrm{Rb}$ ($5^2P_{1/2}$) is well resolved within the 548~MHz Doppler profile~\cite{steck2024}. At the dual-zone temperatures ($75\,^\circ\mathrm{C}$ lower / $85\,^\circ\mathrm{C}$ upper), the resonant optical absorption is approximately 50\%, and the laser frequency is locked continuously to the absorption peak of the $^{87}\mathrm{Rb}$ $F=2 \to F'=1$ transition.

Signal extraction is partitioned between two complementary balanced-polarimeter readout channels (Fig.~\ref{fig:setup}):
\begin{enumerate}
  \item \textbf{Sum Channel ($I_{\mathrm{PD1}} + I_{\mathrm{PD2}}$):} Amplified by an instrumentation amplifier (\texttt{Inst Amp}), representing total optical transmission (absorption). This channel feeds three phase-sensitive detectors demodulated at $f_x = 5460~\mathrm{Hz}$ for the $B_x$ error, at $f_y = 1290~\mathrm{Hz}$ for the $B_y$ error, and at $f_{\mathrm{lock}} = 3530~\mathrm{Hz}$ for the VCSEL direct-absorption frequency lock error.
  \item \textbf{Difference Channel ($I_{\mathrm{PD1}} - I_{\mathrm{PD2}}$):} Detected directly by a low-noise differential transimpedance amplifier (\texttt{I-V Conv}), representing alignment-induced linear dichroism and nonlinear magneto-optical rotation. Because the $45^\circ$ incident polarization balances the two photocurrents at zero field, subtracting $I_{\mathrm{PD1}}$ and $I_{\mathrm{PD2}}$ directly at the virtual-ground transimpedance input cancels quiescent DC photocurrents, common-mode laser relative intensity noise (RIN), and the 3530~Hz laser-dither modulation prior to voltage amplification. Demodulating this high-SNR difference output at $f_z = 2130~\mathrm{Hz}$ extracts the $B_z$ error signal.
\end{enumerate}

\subsection{Control Electronics, Feedback Coils, and Current Sources}
\label{sec:elec}

Orthogonal sinusoidal modulation signals are generated at $f_x = 5460~\mathrm{Hz}$, $f_y = 1290~\mathrm{Hz}$, and $f_z = 2130~\mathrm{Hz}$ (nominal drive amplitude $0.05~\mathrm{V}_{\mathrm{pp}}$), alongside the VCSEL wavelength dither at $f_{\mathrm{lock}} = 3530~\mathrm{Hz}$ ($0.005~\mathrm{V}_{\mathrm{pp}}$). Signal demodulation is performed by four low-cost analog phase-sensitive detector modules based on the AD630 balanced modulator/demodulator~\cite{ad630} (Fig.~\ref{fig:setup}), each equipped with a dedicated second-order LPF ($f_c = 10~\mathrm{Hz}$) that passes geomagnetic variations while rejecting the kHz carriers and 20~kHz heater drive.

Digital PID feedback is executed on an embedded NI myRIO microcontroller operating at a deterministic loop interval of 4~ms (250~Hz update rate). Output records are decimated and logged at 20~Hz (50~ms interval) for spectral noise and dynamic response characterization, and averaged into 1-minute values for multi-day geomagnetic benchmarking. As detailed in Sec.~\ref{sec:comp}, the 12-bit DAC outputs ($\pm 100~\mathrm{nT}$ dynamic window) drive fast compensation currents while analog PI mid-ranging controllers continuously offload slow ambient drifts to three programmable DC constant current sources: a low-noise programmable DC current standard (Yokogawa 7651, operating on its 1~mA range near $-0.34~\mathrm{mA}$ in the unshielded laboratory) for the $x$-axis, and a dual-channel source measure unit (Advantest R6245, operating near $-14.4~\mathrm{mA}$ and $+16.1~\mathrm{mA}$) for the $y$- and $z$-axes.

Throughout all closed-loop experiments reported in this paper, active magnetic field compensation is delivered by an internal three-axis feedback coil set (nominal diameter $\approx 5~\mathrm{cm}$) directly surrounding the insulated copper oven (Fig.~\ref{fig:setup}). The assembly consists of a rigid epoxy-encapsulated 600-turn copper coil pair (inner diameter 40~mm, outer diameter 50~mm) along the optical $x$-axis, combined with compact, low-inductance multi-turn air-core copper coils along the transverse $y$- and $z$-axes. Rather than relying on nominal geometric turn counts for the hand-wound transverse coils, all axial scale factors and non-orthogonalities are established via the external metrological calibration chain described in Sec.~\ref{sec:calib}.

\section{Sensor Calibration and Static/Dynamic Crosstalk Analysis}
\label{sec:calib}

\begin{figure}[htbp]
  \centering
  \includegraphics[width=0.82\linewidth]{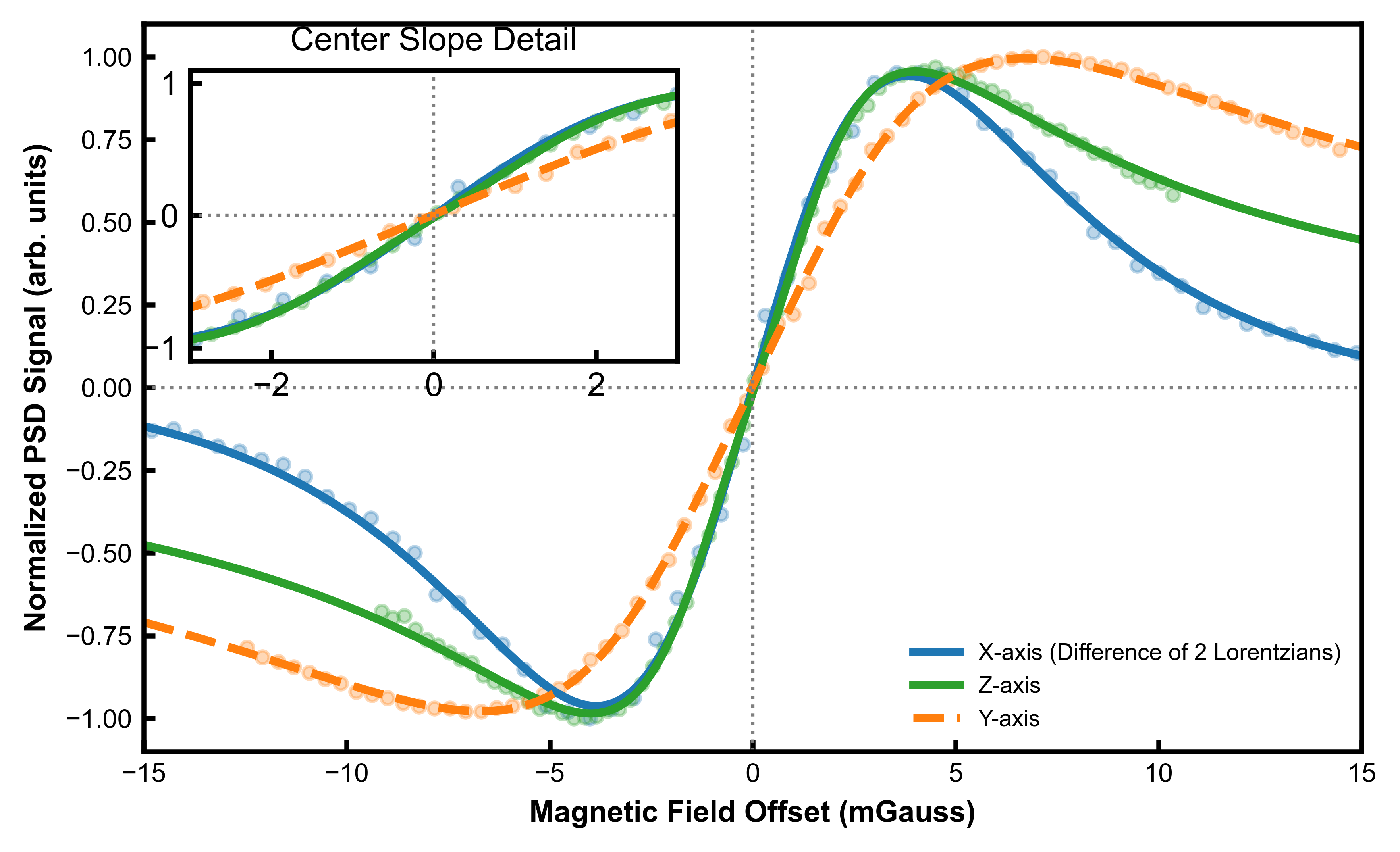}
  \caption{Measured and fitted open-loop magnetic field sweep dispersion curves across the three orthogonal axes around null field. Normalized phase-sensitive detector (PSD) signals are plotted versus magnetic field offset in mGauss ($1~\mathrm{mGauss} = 100~\mathrm{nT}$) for the $B_x$ (blue solid line), $B_y$ (orange dashed line), and $B_z$ (green solid line) axes. The $B_x$ channel exhibits a narrow difference-of-two-Lorentzians profile arising from velocity-selective CPT saturation, and $B_z$ exhibits a similarly narrow $\Delta m_z = 2$ linear dichroism slope, whereas $B_y$ displays roughly twice the linewidth and half the central slope in agreement with Sec.~\ref{sec:hybrid}. Inset: Enlarged view of the central linear region between $-2$ and $+2~\mathrm{mGauss}$ ($\pm 200~\mathrm{nT}$).}
  \label{fig:sweep}
\end{figure}

In single-beam vector atomic magnetometers, mechanical coil misalignments, optical beam projections, and inter-axis quantum-interference couplings must be quantitatively calibrated to achieve accurate three-axis vector reconstruction~\cite{huang2016,zhang2020,li2022}. The normalized open-loop magnetic sweep dispersion curves around zero field for all three axes are presented in Fig.~\ref{fig:sweep}. In direct quantitative agreement with the alignment theory of Sec.~\ref{sec:hybrid}, the $B_x$ and $B_z$ resonances exhibit narrow central features governed by $\Delta m = 2$ coherences (peak-to-peak inflection width $\approx \pm 3.8~\mathrm{mGauss}$ or $\pm 380~\mathrm{nT}$, with a steep linear center slope within $\pm 2~\mathrm{mGauss}$), whereas the $B_y$ absorption resonance is roughly twice as broad ($\approx \pm 7~\mathrm{mGauss}$) with approximately half the central slope.

\subsection{Three-Step Absolute Scale Calibration and Static Decoupling Matrix}
\label{sec:static_cal}

While the internal 5~cm coil assembly serves as the active feedback actuator in all experiments, accurate calibration of its scale factors and static orthogonality was performed using a three-step metrological procedure:
\begin{enumerate}
  \item \textbf{Absolute $x$-Axis Scale Verification via Weak-Field NMR:} The theoretical coil constant of the rigid, epoxy-encapsulated 600-turn $x$-axis coil pair was first calculated from its physical geometry and independently verified against absolute Larmor precession frequency measurements using optically pumped weak-field nuclear magnetic resonance (NMR)~\cite{savukov2005}, tying the $x$-axis field scale directly to fundamental atomic gyromagnetic ratios.
  \item \textbf{Scale Transfer via External 21~cm Helmholtz Coil:} To determine the exact effective coil constants of the hand-wound transverse ($y$- and $z$-axis) feedback coils without relying on turn-count assumptions, an external, high-homogeneity three-axis Helmholtz coil system (diameter 21~cm) was temporarily positioned around the entire sensor head as a symmetric transfer standard. Known uniform calibration currents were applied to the 21~cm coils while the internal 5~cm coils operated in closed-loop zero-field compensation. Cross-calibrating the closed-loop $y$- and $z$-axis feedback currents against the verified $x$-axis reference yielded the accurate absolute scale factors for all three channels.
  \item \textbf{Quasi-DC Static Orthogonality Decoupling:} To isolate intrinsic sensor non-orthogonality from dynamic closed-loop phase lags and reactive eddy currents in the copper heater sleeve, uniform calibration fields were applied sequentially along each axis of the external 21~cm coil at a quasi-DC frequency of 0.01~Hz. Using multivariable least-squares regression between the applied uniform field vector $\bm{B}_{\mathrm{true}}$ and the scale-calibrated raw closed-loop output $\bm{B}_{\mathrm{meas}}$ (setting statistically insignificant entries below 0.005 to zero), the static decoupling matrix $M_{\mathrm{decouple}}$ was obtained:
  \begin{equation}
    \begin{pmatrix} B_{x,\mathrm{true}} \\ B_{y,\mathrm{true}} \\ B_{z,\mathrm{true}} \end{pmatrix}
    =
    \begin{pmatrix}
      1 & 0.050 & 0 \\
      -0.038 & 1 & 0 \\
      0 & 0.067 & 1
    \end{pmatrix}
    \begin{pmatrix} B_{x,\mathrm{meas}} \\ B_{y,\mathrm{meas}} \\ B_{z,\mathrm{meas}} \end{pmatrix}.
    \label{eq:decouple}
  \end{equation}
\end{enumerate}
All static off-diagonal cross-talk elements remained $\le 6.7\%$. Applying $M_{\mathrm{decouple}}$ orthogonalizes the sensor head response to an identity matrix across all three measurement axes.

\subsection{Reconciliation of Static Orthogonality and Dynamic $Z \to Y$ Derivative Coupling}
\label{sec:dyn_cal}

The quantum-interference Jacobian derived in Sec.~\ref{sec:hybrid} ($V_x \propto B_x$, $V_y \propto B_y - B_z$, and $V_z \propto B_z$) resolves a striking asymmetry observed between preliminary self-excitation tests, quasi-DC calibration (0.01~Hz), and higher-frequency dynamic calibration (0.1~Hz).

During preliminary single-coil tests where large perturbations were injected directly through the internal 5~cm $z$-coil---or when the $y$-axis digital PID loop gain was set excessively high (reducing phase margin near the 10~Hz LPF)---an apparent $Z \to Y$ cross-talk approaching 100\% was observed. Two coupled physical effects account for this behavior: first, localized fringe field gradients from the compact 5~cm $z$-coil across the 20~mm active cell length prevented complete zero-field nulling throughout the beam volume; second, as established in Sec.~\ref{sec:hybrid}, any uncompensated residual field $\delta B_z$ enters the $45^\circ$-polarized sum-channel error signal $V_y \propto B_y - \delta B_z$ with an intrinsic 1:1 (100\%) additive weight.

When compensating a uniform external magnetic field (such as the geomagnetic field or the 21~cm calibration coil) at quasi-DC frequencies (0.01~Hz), the independent $z$-axis closed-loop controller continuously drives $\delta B_z \to 0$ across the 3~mm beam volume. Consequently, the intrinsic 1:1 $Z \to Y$ alignment coupling vanishes statically, leaving only the small mechanical non-orthogonality ($\le 6.7\%$) captured by $M_{\mathrm{decouple}}$. Conversely, when the calibration frequency of the 21~cm coil was increased tenfold to 0.1~Hz and evaluated as a complex transfer matrix, a large $90^\circ$ out-of-phase (imaginary) coupling emerged selectively in the $Z \to Y$ channel (reaching $\sim 700\%$ at 0.1~Hz), while the reverse $Y \to Z$ coupling remained small ($\approx 4\%$) and the $x$-axis remained decoupled. This stark directional asymmetry confirms the upper-triangular structure of Sec.~\ref{sec:hybrid}: at 0.1~Hz, the finite integral tracking lag of the $z$-axis feedback loop leaves a transient quadrature residual $\delta B_z \propto j\omega B_{z,\mathrm{ext}}$ inside the vapor cell, which projects directly into the $y$-axis sum channel and is further amplified by the gentler $B_y$ slope and lower damped $y$-axis PID gain. Because the balanced difference channel ($B_z$) is insensitive to first-order $B_y$ perturbations, no reciprocal $Y \to Z$ surge occurs.

From a metrological standpoint, this frequency dependence demonstrates two key principles: (i) static orthogonalization matrices must be calibrated at quasi-DC frequencies ($\le 0.01~\mathrm{Hz}$), because extracting the real part at 0.1~Hz in the presence of a large imaginary term risks severe phase-leakage errors ($\propto \omega^2$); and (ii) while dynamic $Z \to Y$ coupling is negligible across slow diurnal geomagnetic variations ($\sim 10^{-5}\text{--}10^{-3}~\mathrm{Hz}$), transient $y$-axis impulses driven by steep $B_z$ railway spikes can be modeled---and removed in post-processing if needed---via a frequency-proportional complex decoupling matrix $\bm{M}(f) = M_{\mathrm{decouple}} + j (f/f_1) M_{\mathrm{imag}}(f_1)$, equivalent to subtracting a derivative term $\tau_{zy}\, dB_z/dt$ in the time domain.

\section{Synthetic Geomagnetic Reference Model and Analytical Methodology}
\label{sec:synth}

To benchmark continuous unshielded vector measurements in Okayama without a co-located reference geomagnetic observatory, we constructed a synthetic geomagnetic reference model from 1-minute digital records of two primary INTERMAGNET observatories operated by the Japan Meteorological Agency~\cite{love2013,kak2013,kny2013}: Kakioka (KAK; $36.23^\circ\mathrm{N}$, $140.19^\circ\mathrm{E}$) and Kanoya (KNY; $31.42^\circ\mathrm{N}$, $130.88^\circ\mathrm{E}$). The measurement site at Okayama University of Science ($34.69^\circ\mathrm{N}$, $133.93^\circ\mathrm{E}$) lies approximately 592~km southwest of KAK and 462~km northeast of KNY, positioning it nearly along the line connecting the two observatories.

Geomagnetic variations recorded at mid-latitude stations arise primarily from two distinct physical sources: regular solar quiet daily variations ($S_q$) driven by ionospheric wind-dynamo currents fixed relative to local solar time~\cite{yamazaki2017}, and irregular disturbance fields ($\mathrm{Dist}$) driven by solar-wind--magnetosphere coupling that occur quasi-simultaneously on a planetary scale in Universal Time (UT)~\cite{love2009}. To treat these two components consistently during spatial interpolation, each 1-minute observatory time series $B_i(t)$ ($i = x, y, z$ in geographic North--East--Down coordinates) was decomposed using a 180-minute centered rolling mean:
\begin{equation}
  B_{i,S_q}(t) = \mathrm{Rolling~Mean}\!\left(B_i(t),\, 180~\mathrm{min}\right), \qquad
  B_{i,\mathrm{Dist}}(t) = B_i(t) - B_{i,S_q}(t).
  \label{eq:sqdist}
\end{equation}
Because the $S_q$ current vortex tracks the apparent longitudinal motion of the Sun at 4~min per degree of longitude, longitudinal time shifts were applied exclusively to the $S_q$ components to synchronize both observatories to the local solar time of Okayama ($133.93^\circ\mathrm{E}$):
\begin{equation}
  \Delta t_{\mathrm{KAK}} = -25~\mathrm{min} \quad (140.19^\circ\mathrm{E}), \qquad
  \Delta t_{\mathrm{KNY}} = +12~\mathrm{min} \quad (130.88^\circ\mathrm{E}),
  \label{eq:shift}
\end{equation}
while the global disturbance components ($B_{i,\mathrm{Dist}}$) received zero time shift. The synthetic reference vector $\bm{B}_{\mathrm{pred}}(t)$ at Okayama was then calculated using distance-based linear interpolation weights ($w_{\mathrm{KAK}} = 0.41$ and $w_{\mathrm{KNY}} = 0.59$):
\begin{equation}
  B_{i,\mathrm{pred}}(t) = w_{\mathrm{KAK}}\!\left(B_{i,S_q,\mathrm{shifted}}^{\mathrm{KAK}}(t) + B_{i,\mathrm{Dist}}^{\mathrm{KAK}}(t)\right)
  + w_{\mathrm{KNY}}\!\left(B_{i,S_q,\mathrm{shifted}}^{\mathrm{KNY}}(t) + B_{i,\mathrm{Dist}}^{\mathrm{KNY}}(t)\right).
  \label{eq:pred}
\end{equation}

Finally, the 1-minute averaged, statically decoupled output of the magnetometer $\bm{B}_{\mathrm{meas}}(t)$ (Sec.~\ref{sec:static_cal}) was correlated with the geographic reference prediction $\bm{B}_{\mathrm{pred}}(t)$ via multivariable linear regression:
\begin{equation}
  \bm{B}_{\mathrm{meas}}(t) = C\,\bm{B}_{\mathrm{pred}}(t) + \bm{B}_{\mathrm{offset}},
  \label{eq:reg}
\end{equation}
where $C$ is a $3\times 3$ composite environmental transfer matrix capturing laboratory coordinate alignment, local building ferromagnetic distortion, and regional electromagnetic induction, and $\bm{B}_{\mathrm{offset}}$ accounts for the static background magnetization of the laboratory environment.

\section{Unshielded Geomagnetic Observation Results and Transfer Matrix Analysis}
\label{sec:obs}

Continuous unshielded monitoring was conducted inside a ground-floor laboratory at Okayama University of Science over multiple multi-day campaigns, including a continuous seven-day benchmark campaign from September~1 to September~9, 2026 (UTC).

\subsection{Horizontal Tracking Fidelity and Resilience to DC Railway Interference}
\label{sec:tracking}

\begin{figure}[htbp]
  \centering
  \includegraphics[width=0.92\linewidth]{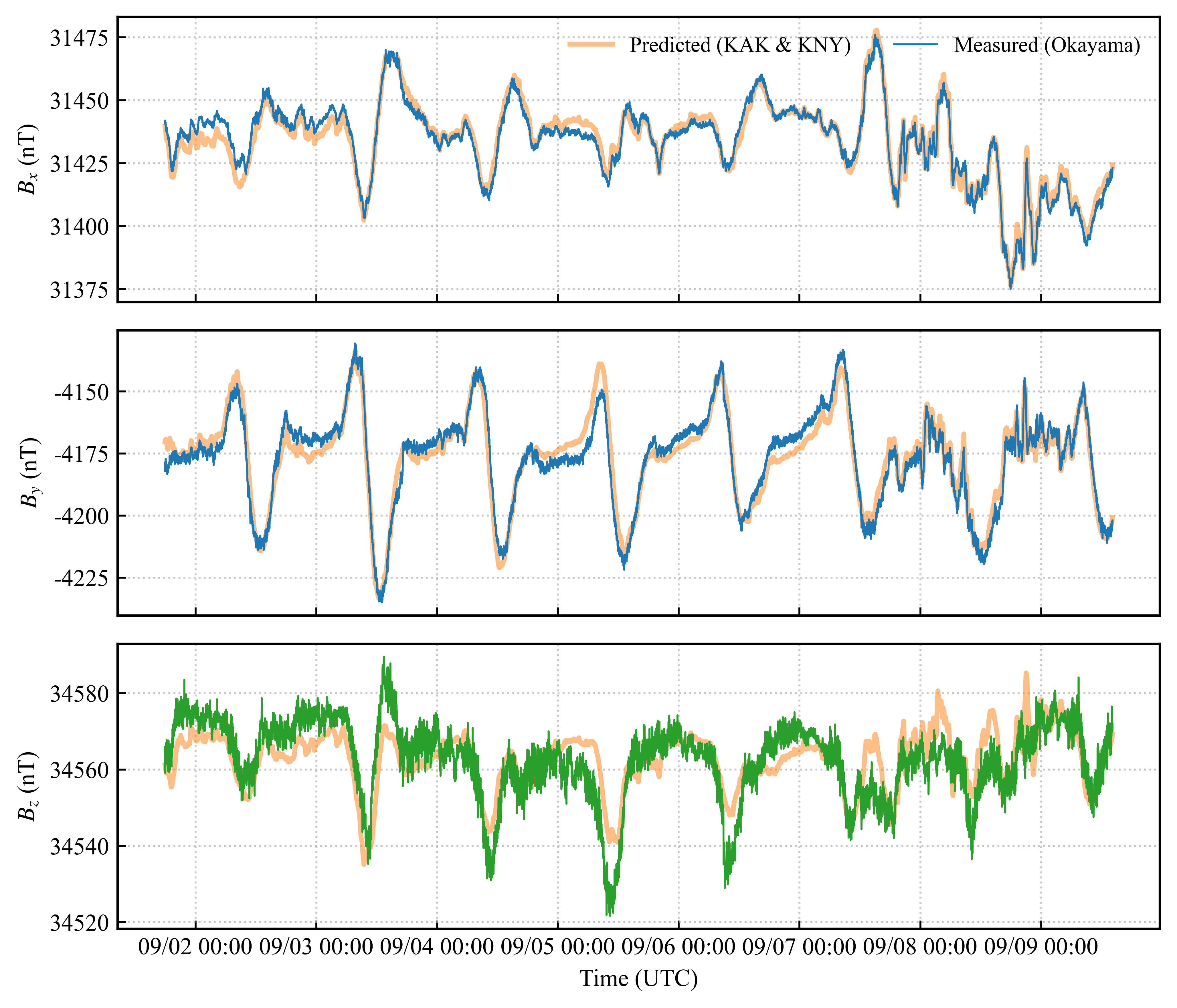}
  \caption{Continuous seven-day geomagnetic tracking under unshielded laboratory conditions in Okayama (September~1--9, 2026 UTC). Traces compare reconstructed Okayama field components ($B_x$ and $B_y$ in blue solid lines, $B_z$ in green solid line; mapped via $C^{-1}$) against synthetic geographic reference predictions (orange semi-transparent lines) derived from Kakioka (KAK) and Kanoya (KNY) 1-minute observatory data via local-time $S_q$ and $\mathrm{Dist}$ decomposition. The horizontal channels closely reproduce diurnal variations and minor pulsations without baseline drift, while $B_z$ tracks daytime high-frequency spikes from regional DC electric railways without closed-loop saturation.}
  \label{fig:tracking}
\end{figure}

Fig.~\ref{fig:tracking} presents the continuous geomagnetic tracking results recorded from September~1 to September~9, 2026 (UTC), comparing the reconstructed vector components (mapped to geographic coordinates via $C^{-1}$) against the KAK/KNY synthetic reference $\bm{B}_{\mathrm{pred}}(t)$. The horizontal components ($B_x$ and $B_y$) exhibit close agreement with the synthetic baseline across the entire multi-day run, accurately reproducing regular diurnal $S_q$ excursions of several tens of nanoteslas as well as minor sub-nanotesla geomagnetic pulsations without noticeable baseline drift.

In contrast, the vertical component ($B_z$) displays persistent high-frequency transient spikes during daytime hours (Fig.~\ref{fig:tracking}). These transients are characteristic of stray ground return currents leaking from regional DC-electrified railway lines (predominantly the San'y\={o} Main Line south of the campus), which form large horizontal ground-current loops that induce prominent transient vertical magnetic fields~\cite{lowes2009}. Notably, in continuous unshielded runs, these transient spikes subside markedly during late-night non-operating railway hours (approximately 00:30--04:45~JST) before resuming with the first morning trains, confirming their anthropogenic origin. Crucially, the dual-stage closed-loop controller tracked continuously through these high-slew railway disturbances without DAC saturation, integrator windup, or frequency-lock dropout, confirming dynamic operational robustness.

\subsection{Polar Decomposition of the Environmental Transfer Matrix: Alignment, Building Distortion, and Regional Induction}
\label{sec:polar}

Least-squares regression of the 1-minute decoupled measurements against the geographic reference $\bm{B}_{\mathrm{pred}}$ over the multi-day campaign yielded the composite transfer matrix $C$ and static offset vector $\bm{B}_{\mathrm{offset}}$:
\begin{equation}
  C = \begin{pmatrix}
    0.938 & -0.164 & 0.053 \\
    0.008 & 1.166 & -0.161 \\
    -0.012 & -0.172 & 1.524
  \end{pmatrix}, \qquad
  \bm{B}_{\mathrm{offset}} = \begin{pmatrix}
    -25022.1 \\ -10789.2 \\ -13745.1
  \end{pmatrix}\mathrm{nT}.
  \label{eq:cmat}
\end{equation}
While the horizontal diagonal scaling factors remain close to unity ($C_{xx} = 0.938$ and $C_{yy} = 1.166$), the vertical diagonal scaling factor is substantially elevated at $C_{zz} = 1.524$ (+52.4\%), accompanied by notable off-diagonal terms in the horizontal plane ($C_{xy} = -0.164$) and between the $y$- and $z$-axes ($C_{yz} = -0.161$ and $C_{zy} = -0.172$). Because the intrinsic scale factors and orthogonality of the magnetometer head were independently calibrated via weak-field NMR and the external 21~cm Helmholtz coil (Sec.~\ref{sec:static_cal}), these matrix deviations originate from the installation geometry and unshielded environment rather than instrumental calibration error.

During installation inside the laboratory, the optical bench was aligned approximately along the building walls and local magnetic meridian rather than surveyed precisely to true geographic North (the mean horizontal field in Fig.~\ref{fig:tracking} corresponds to a magnetic declination of $\arctan(-4190/31430) \approx -7.6^\circ$, consistent with the local declination in Okayama). To rigorously distinguish rigid three-dimensional sensor misalignment from local magnetic field distortion, we performed a polar decomposition $C = S R$, factoring $C$ into a proper orthogonal rotation matrix $R$ ($\det R = +1$) and a symmetric positive-definite stretch tensor $S$:
\begin{equation}
  R = \begin{pmatrix}
    0.9966 & -0.0798 & 0.0209 \\
    0.0797 & 0.9968 & 0.0049 \\
    -0.0212 & -0.0032 & 0.9998
  \end{pmatrix}, \qquad
  S = \begin{pmatrix}
    0.949 & -0.088 & 0.034 \\
    -0.088 & 1.162 & -0.165 \\
    0.034 & -0.165 & 1.524
  \end{pmatrix}.
  \label{eq:polar}
\end{equation}
This decomposition yields three clear physical conclusions:
\begin{enumerate}
  \item \textbf{Rigid-Body Bench Alignment ($R$):} Extracting Euler angles from $R$ reveals a modest horizontal azimuthal yaw of $\approx 4.6^\circ$ relative to true geographic North, with negligible tilt in pitch ($\approx 1.2^\circ$) and roll ($\approx -0.2^\circ$).
  \item \textbf{Local Ferromagnetic Flux Distortion ($S$):} Crucially, the off-diagonal terms $C_{yz} = -0.161$ and $C_{zy} = -0.172$ share the \emph{same} negative sign. A rigid roll tilt of the optical bench about the $x$-axis would instead generate antisymmetric off-diagonal elements of opposite sign ($R_{yz} = -R_{zy}$). Their nearly equal, identically signed values map almost entirely into the symmetric distortion tensor ($S_{yz} = S_{zy} = -0.165$), alongside the pronounced vertical flux amplification ($S_{zz} = 1.524$) and horizontal shear ($S_{xy} = S_{yx} = -0.088$). A symmetric linear distortion tensor of this form is the hallmark of anisotropic magnetic flux concentration and refraction produced by the ferromagnetic steel reinforcement pillars and floor slabs of the laboratory building (consistent with the large static laboratory magnetization $\bm{B}_{\mathrm{offset}}$).
  \item \textbf{Superposition with Regional Geomagnetic Induction (Coast Effect):} In geophysics, electrical conductivity contrasts between the deep Pacific Ocean (adjacent to KAK and KNY), the shallow Seto Inland Sea (adjacent to Okayama), and the resistive continental lithosphere funnel induced telluric currents along coastlines, generating secondary vertical variations ($\Delta B_z = A\,\Delta B_x + B\,\Delta B_y$) characterized by frequency-dependent Parkinson induction vectors~\cite{parkinson1959,parkinson1962,parkinson1979,schmucker1970}. Although this regional coast effect also contributes to the vertical transfer row ($C_{zx} = -0.012, C_{zy} = -0.172$), our polar decomposition demonstrates that inside a steel-reinforced building, local ferromagnetic flux distortion ($S_{yz} = S_{zy} = -0.165, S_{zz} = 1.524$) is superposed with---and dominates over---the regional induction signal. Consequently, while isolating pure Parkinson induction vectors requires outdoor free-field deployment and frequency-domain transfer function analysis, the stationarity of $C$ over multi-day intervals confirms that linear inversion reliably compensates for static indoor ferromagnetic distortion and enables accurate geomagnetic monitoring even within an unshielded laboratory building.
\end{enumerate}

\section{Dynamic Response and Spectral Noise Characterization}
\label{sec:dyn_noise}

\subsection{Dynamic Step Response and Transient Edge Coupling}
\label{sec:step}

\begin{figure}[htbp]
  \centering
  \includegraphics[width=0.90\linewidth]{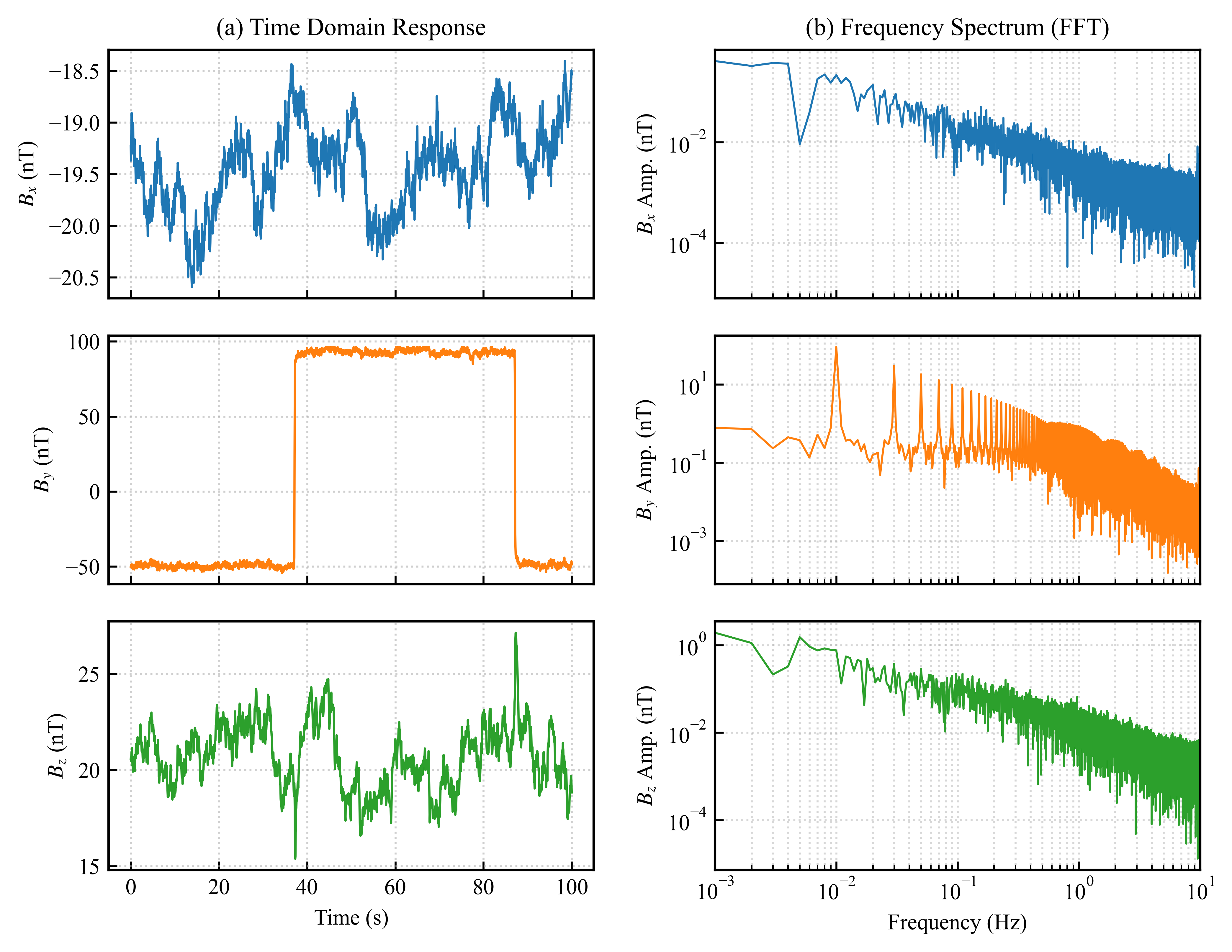}
  \caption{Dynamic response of the three-axis magnetometer to a square-wave magnetic field ($\sim 140~\mathrm{nT}_{\mathrm{p\text{-}p}}$, $\sim 0.01~\mathrm{Hz}$) applied along the $y$-axis via the external 21~cm Helmholtz coil at a 20~Hz sampling rate. (a) Time-domain response (left column): $B_y$ tracks the step transitions cleanly without overshoot or ringing, while $B_x$ and $B_z$ remain free from steady-state DC step cross-talk (minor transient $dB_y/dt$ edge impulses appear briefly on $B_z$ only at the step transitions). (b) Frequency spectra (right column): $B_y$ exhibits distinct odd-harmonic square-wave peaks, with no harmonic leakage above the noise floor in $B_x$ or $B_z$.}
  \label{fig:step}
\end{figure}

To evaluate the dynamic tracking response and axis isolation under calibrated external excitations, a square-wave magnetic field ($\sim 140~\mathrm{nT}_{\mathrm{p\text{-}p}}$, fundamental frequency $\sim 0.01~\mathrm{Hz}$) was applied along the $y$-axis via the external 21~cm Helmholtz coil while recording the statically decoupled closed-loop outputs at a 20~Hz sampling rate. As depicted in Fig.~\ref{fig:step}(a), the $B_y$ time-domain trace tracked the rapid step transitions cleanly without overshoot or ringing, while both $B_x$ and $B_z$ remained completely free from steady-state DC step cross-talk.

At the exact instants of the steep square-wave edges in Fig.~\ref{fig:step}(a), minor transient impulse spikes ($\propto dB_y/dt$) appear briefly on the $B_z$ channel before decaying back to zero. These short-lived edge transients originate from high-slew-rate inductive eddy-current coupling in the surrounding slitted copper heater sleeve and finite closed-loop settling dynamics---a far weaker, localized effect than the intrinsic $Z \to Y$ quantum-interference derivative coupling analyzed in Sec.~\ref{sec:dyn_cal}. In the corresponding frequency spectra (Fig.~\ref{fig:step}(b)), the fundamental square-wave tone ($\sim 0.01~\mathrm{Hz}$) and its odd Fourier harmonics (3rd, 5th, 7th, \dots) stand out clearly in $B_y$, while no harmonic leakage exceeds the baseline noise floor in $B_x$ or $B_z$ across the sub-Hz geomagnetic passband, confirming robust dynamic axis isolation.

\subsection{Baseline Noise Spectra and Sub-LSB Resolution}
\label{sec:noise}

\begin{figure}[htbp]
  \centering
  \includegraphics[width=0.98\linewidth]{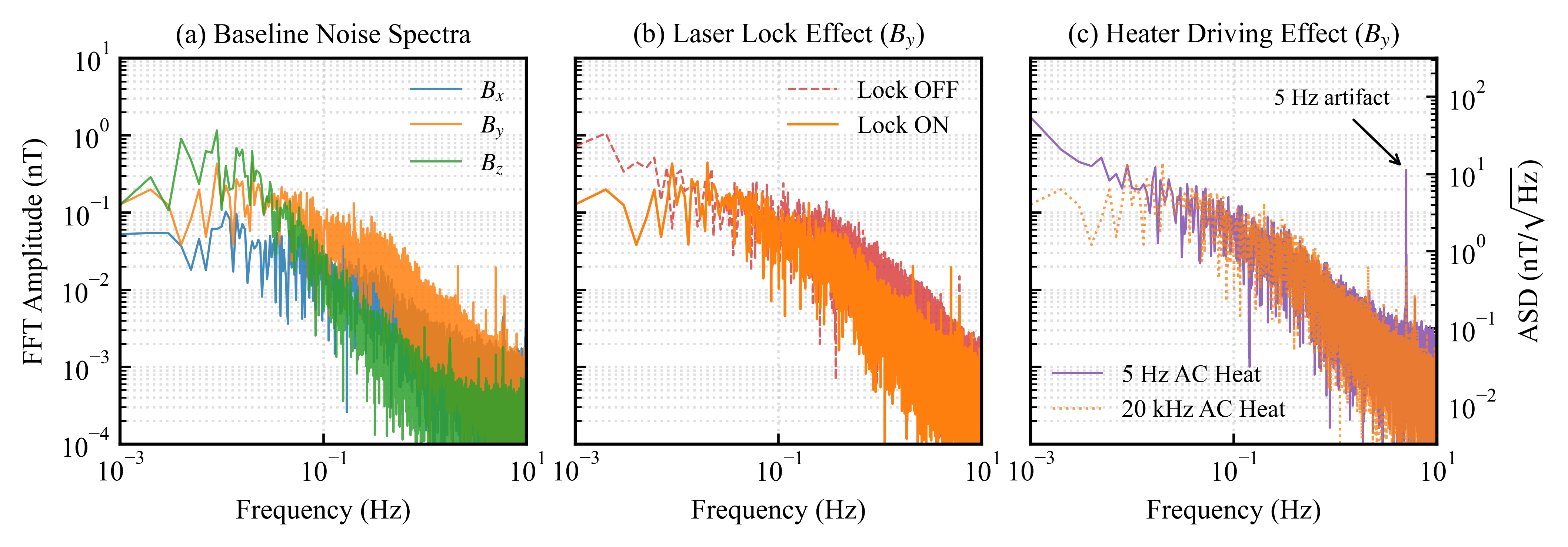}
  \caption{Spectral noise characterization and hardware validation of the closed-loop magnetometer (20,000~points sampled at 20~Hz, bin resolution $\Delta f = 1~\mathrm{mHz}$), displaying both FFT amplitude (left axis, nT) and normalized amplitude spectral density (right axis, $\mathrm{nT}/\sqrt{\mathrm{Hz}}$). (a) Baseline noise spectra of the $B_x$ (blue), $B_y$ (orange), and $B_z$ (green) channels during normal operation, rolling off smoothly above 100~mHz with the 10~Hz LPF and closed-loop bandwidth to reach sub-LSB amplitudes of 0.5--5~pT ($15\text{--}150~\mathrm{pT}/\sqrt{\mathrm{Hz}}$) at 1~Hz and 0.1--1~pT ($3\text{--}30~\mathrm{pT}/\sqrt{\mathrm{Hz}}$) at 10~Hz. (b) Comparison of $B_y$ spectra with the direct-absorption laser frequency lock active (Lock ON, orange solid line) and disabled from resonance (Lock OFF, red dashed line), demonstrating nearly an order-of-magnitude suppression of low-frequency thermal drift below 10~mHz. (c) Effect of heater drive frequency on $B_y$: 5~Hz AC heating introduces a sharp $\sim 0.35~\mathrm{nT}$ artifact at 5~Hz, whereas 20~kHz AC heating completely eliminates the in-band interference.}
  \label{fig:noise}
\end{figure}

Fig.~\ref{fig:noise}(a) presents the spectral noise characterization of the three axes recorded in the unshielded laboratory (20,000~points sampled at 20~Hz, total duration $T = 1{,}000~\mathrm{s}$, frequency bin resolution $\Delta f = 1/T = 1.0~\mathrm{mHz}$). To facilitate direct metrological comparison with other atomic magnetometers, both the raw FFT amplitude in nT (left axis) and the normalized amplitude spectral density ($\mathrm{ASD} = \mathrm{Amplitude}/\sqrt{\Delta f}$, right axis, in $\mathrm{nT}/\sqrt{\mathrm{Hz}}$) are shown. Below 100~mHz, the spectra exhibit a $1/f$-like increase dominated by ambient unshielded geomagnetic fluctuations and programmable current-source flicker noise~\cite{budker2007,schwindt2004}. When evaluating full-bandwidth composite records ($B_{i,\mathrm{total}} = B_{i,\mathrm{CCS}} + B_{i,\mathrm{DAC}}$) containing slow diurnal DC trends, linear detrending combined with Hann-windowed spectral estimation~\cite{welch1967} eliminates end-point spectral leakage across the higher-frequency band, whereas fast-stage DAC records ($B_{i,\mathrm{DAC}}$) exhibit a gentle high-pass roll-off below the mid-ranging crossover ($\sim 10\text{--}30~\mathrm{mHz}$) as slow drifts are offloaded to the CCS.

Above 100~mHz, eliminating the odd-harmonic intermodulation beats (Sec.~\ref{sec:freq}) and tuning the digital PID loop gain removes artificial high-frequency floors, revealing the smooth roll-off governed by the second-order 10~Hz AD630 low-pass filter and closed-loop tracking bandwidth rather than a flat white-noise plateau. At 1~Hz, the $B_x$ and $B_z$ channels reach FFT amplitudes of 0.5--2~pT (corresponding to $\mathrm{ASD} = 15\text{--}60~\mathrm{pT}/\sqrt{\mathrm{Hz}}$), while $B_y$ reaches 3--5~pT ($\sim 100\text{--}150~\mathrm{pT}/\sqrt{\mathrm{Hz}}$). Approaching 10~Hz, $B_y$ falls to $\sim 1~\mathrm{pT}$ ($\sim 30~\mathrm{pT}/\sqrt{\mathrm{Hz}}$), and $B_x$ and $B_z$ extend down to 0.1--0.4~pT ($3\text{--}12~\mathrm{pT}/\sqrt{\mathrm{Hz}}$). The slightly higher noise level on $B_y$ above 100~mHz directly reflects three physical factors identified in Secs.~\ref{sec:hybrid} and~\ref{sec:optics}: its roughly two-fold gentler resonance slope around null field (Fig.~\ref{fig:sweep}), uncancelled common-mode laser relative intensity noise (RIN) in the sum channel, and its lowest modulation carrier frequency (1290~Hz). Further tuning $f_x$ to 5460~Hz and applying stronger loop damping on $B_y$ reduces all three axes to 0.3--2~pT at 1~Hz (0.05--0.2~pT near 10~Hz). Crucially, these 1--10~Hz fluctuation amplitudes lie well below the discrete 12-bit DAC quantization step ($\Delta = 48.8~\mathrm{pT/LSB}$) and the theoretical full-band RMS quantization noise ($\sigma_q = 14.1~\mathrm{pT}$, uniform spectral density $\sigma_q / \sqrt{125~\mathrm{Hz}} \approx 1.26~\mathrm{pT}/\sqrt{\mathrm{Hz}}$), experimentally validating sub-LSB resolution via stochastic dithering and 250~Hz oversampling~\cite{carbone1994}.

\subsection{Low-Frequency Drift Suppression via Laser Frequency Locking}
\label{sec:lock}

Fig.~\ref{fig:noise}(b) compares the $B_y$ noise spectra acquired with the direct-absorption VCSEL frequency lock active (Lock ON, orange solid curve) versus disabled (Lock OFF, red dashed curve). For a fair comparison without artificial slope reduction, the Lock OFF trace was recorded by first tuning the VCSEL precisely to the center of the $^{87}\mathrm{Rb}$ $F=2 \to F'=1$ absorption peak and then releasing the feedback loop to free-run. While the high-frequency roll-off above 50~mHz remains comparable in both states, activating direct-absorption locking suppresses low-frequency baseline fluctuations below 10~mHz by nearly an order of magnitude (from $\sim 1.0~\mathrm{nT}$ down to $\sim 0.13~\mathrm{nT}$ at 1--2~mHz), confirming that active wavelength locking eliminates thermal detuning and FM-to-AM noise conversion on the atomic absorption slope~\cite{affolderbach2005,gruet2013}.

\subsection{Verification of 20 kHz AC Heater Excitation}
\label{sec:heater}

To demonstrate the necessity of ultrasonic AC thermal management, the cell heater was intentionally driven at 5~Hz inside the measurement passband. As shown in Fig.~\ref{fig:noise}(c), the 5~Hz drive generated a sharp interference spike of $\sim 0.35~\mathrm{nT}$ ($\sim 11~\mathrm{nT}/\sqrt{\mathrm{Hz}}$) at exactly 5~Hz across the magnetic spectrum. Switching to 20~kHz AC excitation coupled through series DC-blocking capacitors shifts all heater-induced magnetic fields far beyond the atomic relaxation bandwidth and 10~Hz demodulation filter cutoff, completely eliminating the in-band artifact (Fig.~\ref{fig:noise}(c))~\cite{lu2020}.

\section{Residual Allan Deviation, Thermal Stability, and Uncertainty Budget}
\label{sec:allan}

\begin{figure}[htbp]
  \centering
  \includegraphics[width=0.78\linewidth]{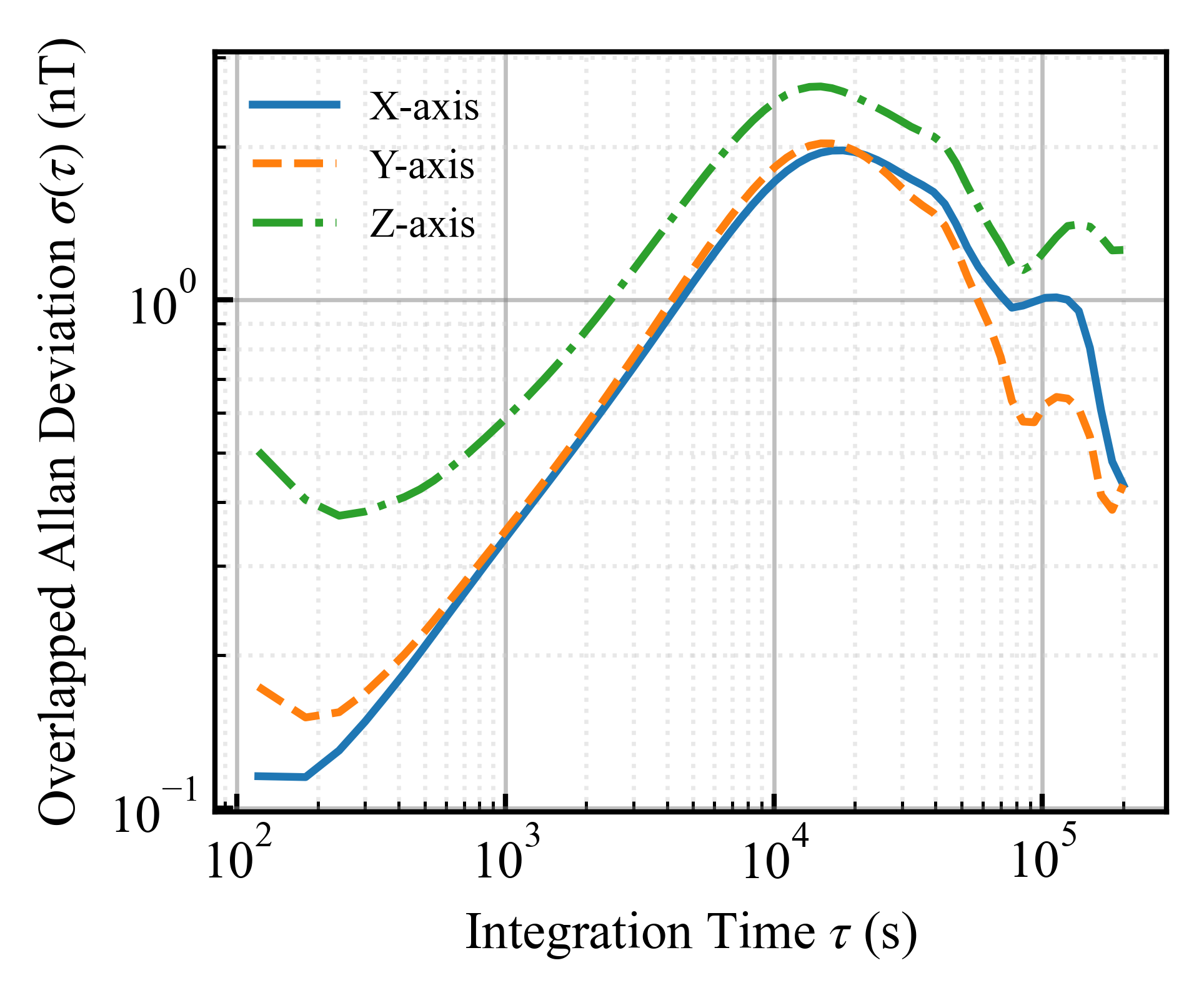}
  \caption{Overlapped Allan deviation $\sigma(\tau)$ calculated from the 1-minute residuals between the measured geomagnetic fields and the synthetic KAK/KNY reference model over the continuous seven-day campaign. At short integration times ($\tau = 60\text{--}200~\mathrm{s}$), the horizontal channels ($B_x$ blue solid line, $B_y$ orange dashed line) establish a conservative upper bound on instrumental bias instability of $\le 0.1~\mathrm{nT}$. Between $\tau = 10^3$ and $2 \times 10^4~\mathrm{s}$, both horizontal axes rise toward 2--3~nT along a closely matched slope, reflecting unmodeled spatial ionospheric heterogeneity combined with diurnal ambient thermal drift of the current sources and copper feedback coils. The elevated level of $B_z$ (green dash-dotted line) reflects regional DC electric railway interference and coastal induction variations.}
  \label{fig:allan}
\end{figure}

To quantify long-term baseline stability unmasked from natural diurnal geomagnetic excursions, the overlapped Allan deviation $\sigma(\tau)$~\cite{allan1966,riley2008} was computed on the 1-minute residuals between the reconstructed Okayama measurements and the KAK/KNY synthetic reference model over the continuous seven-day campaign (Fig.~\ref{fig:allan}).

At the shortest available integration times permitted by the 1-minute observatory cadence ($\tau = 60\text{--}200~\mathrm{s}$), the residual Allan deviation of both horizontal channels ($B_x$ and $B_y$) reaches approximately 0.1~nT (Fig.~\ref{fig:allan}). Because the residual time series inherently includes spatial interpolation errors of the distant reference stations ($\sim 462\text{--}592~\mathrm{km}$ away), this measurement establishes a conservative \emph{upper bound} on the intrinsic instrumental bias instability of $\sigma_{\mathrm{inst}}(\tau) \le 0.1~\mathrm{nT}$ at $\tau = 100\text{--}200~\mathrm{s}$.

For longer integration times ($\tau = 10^3\text{--}2 \times 10^4~\mathrm{s}$, corresponding to timescales of hours to half a day), the horizontal residual Allan deviations rise toward 2--3~nT along a closely matched upward trajectory, while diverging slightly at the shortest and longest $\tau$ (Fig.~\ref{fig:allan}). Meanwhile, the vertical channel ($B_z$) remains elevated across all integration times due to daytime DC railway stray currents~\cite{lowes2009} and coastal induction variations~\cite{parkinson1959,parkinson1962,parkinson1979}.

To rigorously attribute each contribution to the observed short- and long-term uncertainty budget, we evaluated the primary instrumental and environmental error mechanisms summarized in Table~\ref{tab:budget}.

\begin{table}[htbp]
  \centering
  \caption{Quantitative uncertainty and error budget for the unshielded three-axis alignment-based atomic magnetometer.}
  \label{tab:budget}
  \footnotesize
  \begin{tabular}{@{}
    >{\raggedright\arraybackslash}p{0.21\textwidth}
    >{\raggedright\arraybackslash}p{0.27\textwidth}
    >{\raggedright\arraybackslash}p{0.28\textwidth}
    >{\raggedright\arraybackslash}p{0.18\textwidth}@{}}
    \toprule
    \textbf{Error / Uncertainty Source} &
    \textbf{Governing Physical Parameter} &
    \textbf{Estimated Magnetic Contribution} &
    \textbf{Dominant Timescale / Band} \\
    \midrule
    12-bit DAC quantization (oversampled) &
    Step $48.8~\mathrm{pT/LSB}$, 250~Hz dithered &
    $\sim 1.26~\mathrm{pT}/\sqrt{\mathrm{Hz}}$ (0.3--2~pT at 1~Hz) &
    1--10~Hz (sub-LSB regime) \\
    \addlinespace[4pt]
    Programmable CCS flicker noise &
    Yokogawa 7651 (1~mA range, $\sim 3.3~\mathrm{ppm}_{\mathrm{rms}}$); Advantest R6245 (60~mA range) &
    $\sim 0.03\text{--}0.10~\mathrm{nT}_{\mathrm{rms}}$ ($x$); $\sim 0.1\text{--}0.3~\mathrm{nT}_{\mathrm{rms}}$ ($y, z$) &
    10--200~mHz; $\tau = 60\text{--}200~\mathrm{s}$ \\
    \addlinespace[4pt]
    Programmable CCS temperature drift &
    Tempco $\sim 15~\mathrm{ppm/K}$ (7651) to $\sim 100~\mathrm{ppm/K}$ (R6245) &
    $\sim 0.11\text{--}0.47~\mathrm{nT/K}$ ($x$); $\sim 0.5\text{--}1.5~\mathrm{nT/K}$ ($y, z$) &
    Diurnal ($\tau \sim 10^4~\mathrm{s}$) \\
    \addlinespace[4pt]
    Feedback coil thermal expansion &
    Copper $\alpha_{\mathrm{Cu}} \approx 16.5~\mathrm{ppm/K}$ (plus epoxy/winding stress) &
    $\sim 0.8~\mathrm{nT/K}$ at $47~\mu\mathrm{T}$ ($\sim 0.12\text{--}0.22~\mathrm{nT/K}$ at indoor static field) &
    Diurnal ($\tau \sim 10^4~\mathrm{s}$) \\
    \addlinespace[4pt]
    Dual-zone Rb cell temperature error &
    Lower $75.000 \pm 0.007\,^\circ\mathrm{C}$; upper $85.000 \pm 0.017\,^\circ\mathrm{C}$ &
    Negligible ($< 0.05~\mathrm{nT}$; $< 1~\mathrm{nT}$ during $-0.32\,^\circ\mathrm{C}$ room drop) &
    $10^2\text{--}10^5~\mathrm{s}$ \\
    \addlinespace[4pt]
    Ambient geomagnetic / building perturbations &
    KAK/KNY ionospheric heterogeneity; DC railways; adjacent-floor ferromagnetic shifts &
    2--3~nT diurnal residual ($x, y$); 5--15~nT spikes ($z$) &
    Diurnal \& transient \\
    \bottomrule
  \end{tabular}
\end{table}

\begin{enumerate}
  \item \textbf{Current-Source Flicker Limit at $\tau = 60\text{--}200~\mathrm{s}$:} Cancelling a geomagnetic component of $\sim 31.4~\mu\mathrm{T}$ to within $0.1~\mathrm{nT}_{\mathrm{rms}}$ demands a relative current stability of $\sim 3.2~\mathrm{ppm}_{\mathrm{rms}}$. This agrees closely with the low-frequency current noise specification of the Yokogawa 7651 DC standard ($0.02~\mu\mathrm{A}_{\mathrm{p\text{-}p}}$ on the 1~mA range, corresponding to $20~\mathrm{ppm}_{\mathrm{p\text{-}p}} \approx 3.3~\mathrm{ppm}_{\mathrm{rms}}$), while the wider-bandwidth output stage of the Advantest R6245 operating on its 60~mA range explains the slightly higher low-frequency flicker on the $y$- and $z$-axes.
  \item \textbf{Thermal Stability and Diurnal Expansion at $\tau \sim 10^4~\mathrm{s}$:} While the vapor cell itself is regulated to $\pm 7~\mathrm{mK}$ (lower reservoir) and $\pm 17~\mathrm{mK}$ (upper optical zone) inside its 5-mm fluororubber jacket, external hardware remains exposed to ambient room temperature swings. In dedicated multi-day ambient-temperature monitoring runs conducted with the laboratory HVAC and ventilation turned off (eliminating relay-switching and air-draft transients), ambient room temperature remained within a slow diurnal envelope of $\le 0.5\,^\circ\mathrm{C}$. Notably, even during an abrupt $-0.32\,^\circ\mathrm{C}$ ambient temperature drop lasting 30~min, no corresponding baseline step was observed on any magnetic channel relative to the synthetic reference, confirming strong short-term thermal immunity. Nevertheless, over full diurnal cycles ($\tau \sim 1.5\text{--}2 \times 10^4~\mathrm{s}$), uncompensated thermal expansion of the copper feedback coils ($\alpha_{\mathrm{Cu}} \approx 16.5~\mathrm{ppm/K}$, corresponding to $\sim 0.8~\mathrm{nT/K}$ at the full $47~\mu\mathrm{T}$ geomagnetic field, or $\sim 0.12\text{--}0.22~\mathrm{nT/K}$ at the reduced indoor static offsets of 7.5--13.5~$\mu\mathrm{T}$), combined with current-source temperature coefficients ($\sim 0.1\text{--}1.5~\mathrm{nT/K}$) and local building steel thermal stress, acts alongside unmodeled spatial ionospheric heterogeneity between Okayama and KAK/KNY to produce the 2--3~nT peak near $\tau \sim 10^4~\mathrm{s}$. Additionally, in multi-story laboratory buildings, relocation of ferromagnetic furniture on adjacent floors (within 2.5--3.5~m vertically) can introduce static dipole shifts of several nanoteslas, particularly on $B_z$.
\end{enumerate}

\section{Discussion and Future Outlook}
\label{sec:disc}

The results demonstrated here challenge the assumption that long-term stable unshielded vector atomic magnetometers require custom microfabricated buffer-gas vapor cells, sub-Doppler saturated absorption optics, or high-end FPGA lock-in hardware~\cite{kitching2018}. By exploiting ground-state Zeeman quantum interference (rank-2 spin alignment) in an unmodified commercial 1~g rubidium ampoule---paired with 3530~Hz direct-absorption VCSEL locking, dual-zone 20~kHz AC thermal control, and a dual-stage 12-bit mid-ranging architecture---the instrument achieves sub-LSB noise amplitudes (0.5--5~pT at 1~Hz, down to 0.1--1~pT near 10~Hz), condensation-free operation over 29~days, and an upper-bound bias instability of $\le 0.1~\mathrm{nT}$ at $\tau = 100\text{--}200~\mathrm{s}$.

Our quantum-interference analysis (Sec.~\ref{sec:hybrid}) and metrological error budget (Table~\ref{tab:budget}) point directly to three targeted hardware and signal-processing upgrades for future sensor iterations:
\begin{enumerate}
  \item \textbf{Low-Expansion Square Coil Systems:} Replacing the compact circular and hand-wound 5~cm feedback coils with a precision-machined or ceramic/quartz printed-circuit-board (PCB) three-axis square coil system will substantially enlarge the homogeneous field volume relative to the vapor cell and tighten mechanical orthogonality~\cite{crosser2010,merritt1983}. While Crosser \emph{et al.}~\cite{crosser2010} analyzed field-uniformity degradation from geometric imperfections in circular Helmholtz coils, Merritt \emph{et al.}~\cite{merritt1983} demonstrated that multi-coil square geometries yield superior uniform volumes and simplified orthogonal mounting. Fabricating such coils on low-thermal-expansion substrates will simultaneously suppress fringe-gradient artifacts and reduce the $\sim 0.8~\mathrm{nT/K}$ copper thermal expansion drift at diurnal timescales.
  \item \textbf{Second-Harmonic ($2f$) Balanced-Difference Detection for $B_y$:} As established in Secs.~\ref{sec:hybrid} and~\ref{sec:dyn_cal}, extracting $B_y$ from the sum channel at $1f_y$ exposes it to uncancelled common-mode laser RIN and an intrinsic 1:1 additive $Z \to Y$ coupling ($V_y \propto B_y - B_z$). An attractive future refinement is transferring $B_y$ readout to the balanced difference channel using second-harmonic ($2f_y$) phase-sensitive detection~\cite{weis2006,budker2002}. Because $2f$ demodulation corresponds to the second partial derivative $\partial^2 \Delta I_{\mathrm{Diff}} / \partial B_y^2$, it eliminates both common-mode laser RIN and first-order additive $Z \to Y$ offset coupling while rejecting linear background drifts and fundamental modulation feedthrough. However, implementing $2f_y$ detection in unshielded environments entails well-defined practical trade-offs: the maximum theoretical central slope is reduced to $\approx 62\%$ of the $1f$ optimum, a $\sim 1.65\times$ larger modulation amplitude is required (which can induce modulation broadening on $B_x$ and $B_z$), and the outer zero-crossing of the second-derivative lineshape ($\sim 2.1$ resonance half-widths from null) introduces a feedback sign reversal that narrows the capture range against abrupt railway or ferromagnetic spikes. Evaluating whether the RIN rejection of the difference channel outweighs these capture-range constraints using a low-distortion digital lock-in stage remains a promising direction for next-generation alignment magnetometers.
  \item \textbf{Ultra-Low-Noise Fixed-Bias Current Offloading:} Because Table~\ref{tab:budget} identifies programmable current-source flicker and temperature drift as the primary instrumental limit at $\tau = 60\text{--}10^4~\mathrm{s}$, offloading $\sim 90\%$ of the static background field onto a dedicated fixed-current stage built from an ovenized buried-Zener voltage reference ($\sim 0.2~\mathrm{ppm/K}$, $\sim 0.13~\mathrm{ppm}_{\mathrm{p\text{-}p}}$ noise), zero-drift chopper operational amplifiers, and ultra-low-TCR bulk metal-foil sense resistors ($\le 0.2~\mathrm{ppm/K}$) will allow the programmable sources to operate on 1--2 orders lower current ranges, potentially lowering the instrumental bias instability well below 0.1~nT.
\end{enumerate}

With its low SWaP-C footprint, resilience against railway interference, and demonstrated vector discrimination of nearby ferromagnetic dipole motions (such as distinguishing opposite polarity signatures of separate laboratory doors), this alignment-based platform offers a practical path toward dense geomagnetic monitoring arrays, volcanological and tectonic sensor networks, and mobile magnetic anomaly detection (MAD).

\section{Conclusion}
\label{sec:conc}

We have designed, modeled, calibrated, and validated a three-axis single-beam alignment-based atomic magnetometer operating continuously in an unshielded geomagnetic environment. Built around an unmodified, uncoated, buffer-gas-free commercial rubidium ampoule and a single 795~nm VCSEL with $45^\circ$-oriented linear polarization, the sensor leverages ground-state Zeeman quantum interference (rank-2 spin alignment) to cancel first-order vector light shifts intrinsically. Dual-zone 20~kHz AC heating ($75.0\,^\circ\mathrm{C}$ lower reservoir / $85.0\,^\circ\mathrm{C}$ upper optical zone) inside a fluororubber-insulated slitted copper sleeve prevented window condensation over 29~days of continuous operation while eliminating in-band thermal magnetic artifacts. By optimizing mutually non-harmonic modulation frequencies ($f_x = 5460~\mathrm{Hz}$, $f_y = 1290~\mathrm{Hz}$, $f_z = 2130~\mathrm{Hz}$, and $f_{\mathrm{lock}} = 3530~\mathrm{Hz}$) within a hybrid balanced polarimeter (\texttt{Inst Amp} sum channel for $B_x, B_y$, and DAS locking; differential \texttt{I-V Conv} channel for $B_z$) and a 12-bit dual-stage mid-ranging closed loop, we achieved sub-LSB noise amplitudes of 0.5--5~pT ($15\text{--}150~\mathrm{pT}/\sqrt{\mathrm{Hz}}$) at 1~Hz rolling off to 0.1--1~pT ($3\text{--}30~\mathrm{pT}/\sqrt{\mathrm{Hz}}$) near 10~Hz. Absolute scale calibration via weak-field NMR and an external 21~cm Helmholtz coil decoupled static cross-axis errors to $\le 6.7\%$ and clarified the quantum-interference origin of dynamic $Z \to Y$ derivative coupling. Continuous seven-day unshielded geomagnetic monitoring in Okayama confirmed high horizontal tracking fidelity against synthesized Kakioka and Kanoya observatory records, separated rigid bench rotation from indoor ferromagnetic distortion via polar decomposition, and established an upper bound on instrumental bias instability of $\le 0.1~\mathrm{nT}$ at integration times of 100--200~s.

\section*{Acknowledgments}
The geomagnetic 1-minute digital data from the Kakioka (KAK) and Kanoya (KNY) observatories were provided by the Kakioka Magnetic Observatory, Japan Meteorological Agency~\cite{kak2013,kny2013}. The author also gratefully acknowledges an anonymous reviewer of an earlier version of this manuscript for insightful comments regarding odd-harmonic intermodulation and thermal error-budget analysis, which directly guided the frequency optimization and quantitative uncertainty analysis reported in this work.

\section*{Data Availability Statement}
The data that support the findings of this study are available from the corresponding author upon reasonable request.

\section*{Conflict of Interest}
The author declares no conflict of interest.

\section*{Ethical Statement}
This study did not involve human participants or animal subjects.

\end{document}